\documentclass[10pt]{article}
\usepackage[latin9]{inputenc}
\pdfoutput=1  

\usepackage{amsmath}
\usepackage{amssymb}
\usepackage{graphicx}

\makeatletter

\providecommand{\tabularnewline}{\\}

\usepackage{enumitem}		

\usepackage{M12}

\definecolor{midblue}{rgb}{0.55, 0.55, 0.9}

\hypersetup{citecolor={darkblue},urlcolor={darkblue} }

\titlelabel{\makebox[7mm][l]{\thetitle}}

\renewcommand{\vec}[1]{\oldvec{#1}\,}

\renewcommand{\hat}[1]{\oldhat{#1}\,}

\usepackage{upgreek}
\renewcommand{\phi}{\upphi}
\renewcommand{\psi}{\uppsi}
\renewcommand{\omega}{\upomega}

\AtBeginDocument{
  
}

\makeatother

\begin{document}
\newcommand{\bose}{%
\begin{tikzpicture}[scale=0.4] 
  \draw [fill=white] (0,0) circle (2mm); 
\end{tikzpicture}}

\newcommand{\boseone}{%
\begin{tikzpicture}[scale=0.4] 
  \draw [fill=newblue] (0,0) circle (2mm); 
\end{tikzpicture}}

\newcommand{\bosetwo}{%
\begin{tikzpicture}[scale=0.4] 
  \draw [fill=neworange] (0,0) circle (2mm); 
\end{tikzpicture}}

\newcommand{\fermi}{%
\begin{tikzpicture}[scale=0.4] \draw (0,0) circle (2mm);
  \draw [thin] (-2mm,0) -- (2mm,0);
  \draw [thin] (0,-2mm) -- (0,2mm);
\end{tikzpicture}}

\newcommand{\fermione}{%
\begin{tikzpicture}[scale=0.4] \draw [fill=newblue] (0,0) circle (2mm);
  \draw [thin] (-2mm,0) -- (2mm,0);
  \draw [thin] (0,-2mm) -- (0,2mm);
\end{tikzpicture}}


\definecolor{newblue}{rgb}{0.7, 0.7, 1}
\definecolor{neworange}{rgb}{0.7, 0.3, 0}
\newcommand{\colourone}{newblue} \newcommand{\colourtwo}{neworange} 

\renewcommand{\address}[2]{ \texorpdfstring{ \small\itshape\smallskip \\ \llap{\makebox[3mm][l]{\textsuperscript{\upshape\scriptsize#1}}}#2 }{} } \settitlesize{20.4pt} 

\title{\ \\ \textmd{Macroscopic (and Microscopic) Massless Modes}}

\author{Michael C. Abbott\alabel{\ \bose} and\hspace{1mm} In\^{e}s Aniceto\alabel{\ \fermi}
\address{\boseone}{QGASLAB, Department of Mathematics, University
of Cape Town,\\
Rondebosch 7701, Cape Town, South Africa.} \address{\fermi}{CAMGSD,
Departamento de Matem\'{a}tica, Instituto Superior T\'{e}cnico,
\\
Av. Rovisco Pais, 1049-001 Lisboa, Portugal.} \address{\fermione}{Institute
of Physics, Jagiellonian University, \\
Ul. \L{}ojasiewicza 11, 30-348 Krak\'{o}w, Poland.} \address{ }{michael.abbott@uct.ac.za,
ianiceto@math.ist.utl.pt} }

\date{19 December 2014\\
arXiv:1412.6380}
\maketitle
\begin{abstract}
We study certain spinning strings exploring the flat directions of
$AdS_{3}\times S^{3}\times S^{3}\times S^{1}$, the massless sector
cousins of $su(2)$ and $sl(2)$ sector spinning strings. We describe
these, and their vibrational modes, using the $D(2,1;\alpha)^{2}$
algebraic curve. By exploiting a discrete symmetry of this structure
which reverses the direction of motion on the spheres, and alters
the masses of the fermionic modes $s\to\kappa-s$, we find out how
to treat the massless fermions which were previously missing from
this formalism. We show that folded strings behave as a special case
of circular strings, in a sense which includes their mode frequencies,
and we are able to recover this fact in the worldsheet formalism.
We use these frequencies to calculate one-loop corrections to the
energy, with a version of the Beisert--Tseytlin resummation. 
\end{abstract}
\tableofcontents{}

\newcommand{\cJ}{\scalebox{0.95}{$\mathcal{J}$}}
\newcommand{\cS}{\scalebox{0.95}{$\mathcal{S}$}}

\newcommand{\cJh}{\cJ_{\!\! h}}
\newcommand{\cSn}{\cS_{\nu}}

\pagebreak{}

\section{Introduction}

One of the new features of the integrable AdS$_{3}$/CFT$_{2}$ correspondence
is the presence of massless modes in the BMN spectrum \cite{Babichenko:2009dk}.%
\footnote{We discuss only the string theory side; for the dual theory see \cite{Gukov:2004ym,Tong:2014yna},
and \cite{Pakman:2009mi,Sax:2014mea} for the related $AdS_{3}\times S^{3}\times T^{4}$
case.%
} Each mode corresponds to a direction away from the string's lightlike
trajectory, and its mass is related to the radius of curvature of
in this direction. In $AdS_{5}\times S^{5}$ all radii and all masses
are equal, but in $AdS_{4}\times CP^{3}$ there are two distinct radii,
and $AdS_{3}\times S^{3}\times S^{3}\times S^{1}$ four: the two 3-spheres
are $1/\cos\phi$ and $1/\sin\phi$ times the $AdS$ radius (where
$\phi$ is an adjustable parameter). The flat $S^{1}$ direction,
$u$, gives one massless boson, but the more interesting one arises
from a combination of the equators of the two $S^{3}$ factors, $\psi$. 

While there is no particular difficulty about treating the massless
modes in the worldsheet language \cite{Sundin:2013ypa,Roiban:2014cia},
they have until recently been missing from the integrable description:
they do not appear in the Bethe equations of \cite{Borsato:2012ss},
the S-matrix of \cite{Borsato:2012ud} and the unitarity methods of
\cite{Bianchi:2013nra,Bianchi:2014rfa}, nor the coset description
of \cite{Zarembo:2010yz}. Recent progress on this problem (and the
related one in $AdS_{3}\times S^{3}\times T^{4}$) has been reported
in \cite{Sax:2012jv,Lloyd:2013wza,Borsato:2014exa,Borsato:2014hja},
and we build on the work of Lloyd and Stefa\'{n}ski \cite{Lloyd:2013wza}
who studied the problem of how to incorporate the $\psi$ direction
in the algebraic curve. This formalism maps the string to a Riemann
surface, given by the log of the eigenvalues of the monodromy matrix
\cite{Kazakov:2004qf,Kazakov:2004nh,SchaferNameki:2004ik,Beisert:2005bm}.
And what they showed is that the traditional way in which the Virasoro
constraint was imposed here (as a condition on the residues of the
quasimomenta at $x=\pm1$) is too strong: it is not implied by the
worldsheet Virasoro constraint when the target space has more than
two factors. By loosening this restriction they were able to study
some classical string solutions (with oscillatory $\psi$) which were
previously illegal. 

What the algebraic curve formalism has proven extremely useful for,
and what we will use it for here, is working out the frequencies of
vibrational modes (especially fermionic modes) of various macroscopic
classical string solutions. The most important examples have been
circular \cite{Frolov:2003qc,Arutyunov:2003za,Park:2005ji,Beisert:2005cw,Hernandez:2006tk,Gromov:2007aq,Mikhaylov:2010ib,Beccaria:2012kb}
and folded \cite{deVega:1996mv,Gubser:2002tv,Frolov:2002av,McLoughlin:2008he,Gromov:2011bz,LopezArcos:2012gb,Forini:2012bb,Bianchi:2014ada}
spinning strings,%
\footnote{Our references here are far from exhaustive, see reviews \cite{Tseytlin:2010jv,McLoughlin:2010jw}
for more.%
} described by one- or two-cut resolvents. The frequencies of these
modes can be added up to give the one-loop correction to the energy,
and this can be efficiently done using the algebraic curve \cite{SchaferNameki:2006gk,Gromov:2008ec,Abbott:2010yb,Gromov:2011bz}.
The $D(2,1;\alpha)^{2}$ algebraic curve needed here has been studied
in \cite{Babichenko:2009dk,Abbott:2012dd,Abbott:2013mpa,Lloyd:2013wza},%
\footnote{Here $\alpha=\cos^{2}\phi$; at $\alpha=1/2$ the algebra becomes
\emph{osp}(4\textbar{}2), for which \cite{Zarembo:2010yz} describes
the algebraic curve in detail. %
} and the computation of $6+6$ mode frequencies (corresponding to
the massive BMN modes) can be done using well-worn tools. 

In the integrable picture, the point-particle BMN state is the ferromagnetic
vacuum of the spin chain, and modes are single impurities (or magnons).
Macroscopic classical solutions are usually%
\footnote{The exception is the giant magnon \cite{Hofman:2006xt}; it is unclear
what if anything the massless analogue of this should be.%
} condensates of a very large number of impurities, arranged such that
they form a single Bethe string in the case of a circular string,
or two in the case of a folded string. Since all the impurities are
of the same type, we need only a small sector of the full Bethe equations:
the $su(2)$ sector for strings exploring $S^{3}$, or $sl(2)$ for
strings in $AdS_{3}$. The energy of a solution to these equations
can be compared to that from a semiclassical calculation of the type
mentioned above, and such comparisons provided important tests of
our understanding of integrable AdS/CFT. 

In this paper we study some classical string solutions which explore
the flat directions of $AdS_{3}\times S^{3}\times S^{3}\times S^{1}$:
a circular string and a folded spinning string. These may perhaps
be thought of as macroscopic condensates of the massless modes, in
the same sense that spinning strings in $S^{3}$ are condensates of
massive modes. The solutions are very simple indeed, since they explore
only a flat torus, but as far as we know have not been viewed in this
light before. We study them and their modes in both the worldsheet
sigma-model language and using algebraic curves, and are able to show
agreement between these two. Some interesting features of this are:
\begin{itemize}
\item The folded string behaves as a special case of the circular string.
In the algebraic curve both are now described by poles alone, and
the folded string simply has no winding, thus equal residues at $x=\pm1$.
This may point to the degeneration of the distinction between one-cut
and two-cut solutions. In the worldsheet picture this agreement is
less obvious, and the modes are most naturally written in a rather
strange gauge. 
\item Those fermions which are massless for the BMN solution are no longer
massless here, and we show how to calculate their frequencies from
the algebraic curve by extending the previously understood formalism
to include all $8$ fermions. To do this we exploit the symmetry that
reverses the direction in which the BMN solution moves, which re-arranges
the fermion masses%
\footnote{We write $\mbox{mass}=s$ throughout, to reserve $m$ for winding
numbers. The heaviest modes (those in $AdS$ directions) have mass
$s=\kappa$.%
} $s\to\kappa-s$. Because we can perform this reversal continuously,
we can follow the behaviour of individual modes as they become massless. 
\item By approaching the BMN solution, we learn something about the massless
limit. The microscopic cuts which describe a vibrational mode approach
the poles at $x=\pm1$ as $s\to0$, while their energies and mode
numbers remain finite. This appears to be analogous to what happens
to the macroscopic classical solutions, for which the cuts of their
massive cousins have been replaced by just poles. 
\item The mode frequencies (for the folded or circular string) give a logarithmic
divergence when naively applying Beisert and Tseytlin's method of
calculating $\delta E$ as an expansion in $1/\cJ$ \cite{Beisert:2005cw}.
However we are able to modify this procedure to give a finite answer. 
\end{itemize}

\subsection*{Outline}

In section \ref{sec:Worldsheet-Sigma-Model} we set up the classical
solutions to be studied, and then calculate their mode frequencies
from the Polyakov and Green--Schwarz actions. In section \ref{sec:Algebraic-Curve}
we find the corresponding algebraic curves, and use the comparison
to guide us to an understanding of how to calculate the previously
missing fermionic modes in $\smash{AdS_{3}\times S^{3}\times S^{3}\times S^{1}}$.
Section \ref{sec:Energy-Corrections} uses these frequencies to compute
compute energy corrections, by adapting the Beisert--Tseytlin re-summing
procedure. Section \ref{sec:Conclusions} concludes. 

Appendix \ref{sec:Reversal-Symmetry-and-AdS4xCP3} looks at the same
reversal symmetry in $AdS_{4}\times CP^{3}$, as a check of our understanding.

\section{Worldsheet Sigma-Model\label{sec:Worldsheet-Sigma-Model}}

We study two kinds of extended string solutions. The circular (or
spiral) spinning string explores a torus $S^{1}\times S^{1}$, stretching
along one diagonal of this square and moving along the other: 
\begin{equation}
\phi_{1}=\omega\tau+m\sigma,\qquad\phi_{2}=\omega\tau-m\sigma.\label{eq:toy-circular}
\end{equation}
These two angles are often taken to be on the same $S^{3}$, with
$ds^{2}=d\theta^{2}+\sin^{2}\theta\, d\phi_{1}^{2}+\cos^{2}\theta\, d\phi_{2}^{2}$.
But the solution can exist in flat space, or in $AdS_{3}\times S^{1}$
with one circle being $\rho=\mbox{const.}$ in $AdS$ \cite{Arutyunov:2003za}.
The folded spinning string instead explores a 2-disk, 
\begin{equation}
X+iY=e^{i\nu\tau}f(\sigma).\label{eq:toy-folded}
\end{equation}
In flat space $f(\sigma)=\smash{\tfrac{1}{\nu}}\cos\nu\sigma$; this
is the string which gives rise to Regge trajectories \cite{Lund:1976ze}.
In curved spaces this becomes one of GKP's solutions \cite{Gubser:2002tv}.
The turning points of $f(\sigma)$ are cusps, at which the induced
metric will have a curvature singularity \cite{Frolov:2002av}. 

We will study these solutions in $AdS_{3}\times S^{3}\times S^{3}\times S^{1}$,
for which the metric is 
\begin{align}
ds^{2} & =R^{2}\left[ds_{AdS}^{2}+\frac{1}{\cos^{2}\phi}ds_{S+}^{2}+\frac{1}{\sin^{2}\phi}ds_{S-}^{2}+du^{2}\right]\displaybreak[0]\label{eq:metric-ads3s3s3s1}\\
\shortintertext{with}ds_{S\pm}^{2} & =d\theta_{\pm}^{2}+\sin^{2}\theta_{\pm}d\varphi_{\pm}^{2}+\cos^{2}\theta_{\pm}d\beta_{\pm}^{2}\nonumber \\
ds_{AdS}^{2} & =-\cosh^{2}\rho\, dt^{2}+d\rho^{2}+\sinh^{2}\rho\, d\gamma^{2}.\nonumber 
\end{align}
Defining $\varphi$ and $\psi$ by%
\footnote{Thus $\varphi_{+}=\cos^{2}\phi\:\varphi-\cos\phi\sin\phi\:\psi$ and
$\varphi_{-}=\sin^{2}\phi\:\varphi+\cos\phi\sin\phi\:\psi$. Our notation
follows \cite{Forini:2012bb} mostly.%
} 
\[
\varphi=\varphi_{+}+\varphi_{-},\qquad\smash{\psi=-\tan\phi\:\varphi_{+}+\cot\phi\:\varphi_{-}}
\]
the metric near to $\rho=0$, $\theta_{\pm}=\frac{\pi}{2}$ is  
\[
ds^{2}/R^{2}=-dt^{2}+d\varphi^{2}\;+d\psi^{2}+du^{2}\;+\bigo{\theta_{\pm}-\tfrac{\pi}{2}}^{2}+\bigo{\rho}^{2}.
\]
The classical solutions we will study here explore only the ``massless''
directions $\varphi,\psi$ and $u$, thus live in $\mathbb{R}\times(S^{1})^{3}$.
The bosonic action (in conformal gauge) is 
\[
S=R^{2}\int\frac{d\tau d\sigma}{4\pi}\eta^{\mu\nu}\partial_{\mu}X^{M}\partial_{\nu}X^{N}G_{MN},\qquad R^{2}=\sqrt{\lambda}=4\pi g
\]
where we write the metric scaled as $ds^{2}=R^{2}\, dX^{M}dX^{N}G_{MN}$.
Since $G_{MN}$ is block-diagonal, the resulting equations of motion
treat each factor in $AdS_{3}\times S^{3}\times S^{3}\times S^{1}$
independently; they are coupled only by the Virasoro constraints.
We write the diagonal and off-diagonal constraints as follows: 
\begin{align}
0=\partial_{0}X^{M}\partial_{0}X^{N}G_{MN}+\partial_{1}X^{M}\partial_{1}X^{N}G_{MN} & =V_{AdS}^{\text{diag}}+V_{S+}^{\text{diag}}+V_{S-}^{\text{diag}}+V_{u}^{\text{diag}}\label{eq:virasoro-terms-WS}\\
0=\partial_{0}X^{M}\partial_{1}X^{N}G_{MN} & =V_{AdS}^{\text{off}}+\ldots\mbox{ similar. }\nonumber 
\end{align}
The solutions usually studied consist of a known solution in each
factor (such as a giant magnon \cite{Abbott:2012dd} or a circular
string \cite{Beccaria:2012kb}) each contributing a constant to the
constraint. Any nontrivially new solution must break this, and we
will study some solutions which do so below. This is a novel feature
also explored (in different language) by \cite{Lloyd:2013wza}. 

We will be concerned with only one Noether charge from each factor
of the space, namely 
\begin{align*}
\Delta & =R^{2}\int\frac{d\sigma}{2\pi}\cosh^{2}\rho\:\partial_{\tau}t, & J_{+} & =\frac{R^{2}}{\cos^{2}\phi}\int\frac{d\sigma}{2\pi}\sin^{2}\theta_{+}\:\partial_{\tau}\varphi_{+}\displaybreak[0]\\
J_{-}= & \frac{R^{2}}{\sin^{2}\phi}\int\frac{d\sigma}{2\pi}\sin^{2}\theta_{-}\:\partial_{\tau}\varphi_{-}, & J_{u} & =R^{2}\int\frac{d\sigma}{2\pi}\:\partial_{\tau}u\:.
\end{align*}
We will also need the total $J'=J_{\varphi}=\cos^{2}\phi\: J_{X}+\sin^{2}\phi\: J_{Y}$,
and will use variants $\cJ=J'/\sqrt{\lambda}$ etc. and $\kappa=\Delta/\sqrt{\lambda}$
.

\subsection{Classical Solutions}

Here are the solutions we study; they are also drawn in figure \ref{fig:classical-solns-2-3-5}:
\begin{enumerate}
\item The supersymmetric BMN ``vacuum'' solution is $\varphi=t=\kappa\tau$,
$\psi=0$, a point particle \cite{Berenstein:2002jq,Babichenko:2009dk}.
The non-supersymmetric vacuum is the generalisation to $\zeta\neq\phi$
in 
\begin{equation}
\varphi_{+}=\kappa\cos\phi\cos\zeta\:\tau,\qquad\varphi_{-}=\kappa\sin\phi\sin\zeta\:\tau,\qquad t=\kappa\tau.\label{eq:vacuum-zeta-class-WS}
\end{equation}
Note that while $\phi\in[0,\tfrac{\pi}{2}]$ is sufficient to describe
all $\cos^{2}\phi=\alpha\in[0,1]$, we will allow $\zeta\in[0,2\pi]$
so that this solution can move in either direction on the $\varphi_{\pm}$
circles; for this reason we do not use $\delta=\cos^{2}\zeta$. The
charges are 
\begin{equation}
\Delta=4\pi g\:\kappa,\qquad J_{+}=4\pi g\:\kappa\:\frac{\cos\zeta}{\cos\phi},\qquad J_{-}=4\pi g\:\kappa\:\frac{\sin\zeta}{\sin\phi}.\label{eq:vacuum-charges-WS}
\end{equation}

\item The most general circular string within the equators of $S^{3}\times S^{3}\times S^{1}$
is 
\begin{equation}
\varphi_{\pm}=\omega_{\pm}\tau+m_{\pm}\sigma,\qquad u=\varphi_{\pm}=\omega_{u}\tau+m_{u}\sigma,\qquad t=\kappa\tau+m_{0}\sigma.\label{eq:circular-class-WS}
\end{equation}
For a closed string we must have $m_{\pm},m_{u}\in\mathbb{Z}$. Physically
$m_{0}=0$, but we include it temporarily in order to allow fluctuations
$\delta m_{0}\neq0$ later. The Virasoro constraints impose
\begin{equation}
\begin{gathered}V^{\text{diag}}=(-\kappa^{2}-m_{0}^{2})+\frac{\omega_{+}^{2}+m_{+}^{2}}{\cos^{2}\phi}+\frac{\omega_{-}^{2}+m_{-}^{2}}{\sin^{2}\phi}+(\omega_{u}^{2}+m_{u}^{2})=0\\
V^{\text{off}}=-\kappa m_{0}+\frac{\omega_{+}m_{+}}{\cos^{2}\phi}\:+\:\frac{\omega_{-}m_{-}}{\sin^{2}\phi}\:+\:\omega_{u}m_{u}\:=0.
\end{gathered}
\label{eq:circular-WS-vir}
\end{equation}
 Each term here comes from one factor of the space, and each is a
constant. The simplest case is to demand that there is no winding
along the $\varphi$ direction, and no momentum in the $\psi$ direction,
giving 
\begin{equation}
\omega_{+}=\cos^{2}\phi\:\omega,\qquad\omega_{-}=\sin^{2}\phi\:\omega,\qquad m_{\pm}=\pm m,\qquad\omega_{u}=m_{u}=0.\label{eq:circular-simple-case}
\end{equation}
The Virasoro constraint then reads $\kappa^{2}=\omega^{2}+4m^{2}/\sin^{2}2\phi$.
\begin{figure}
\centering \includegraphics[width=45mm]{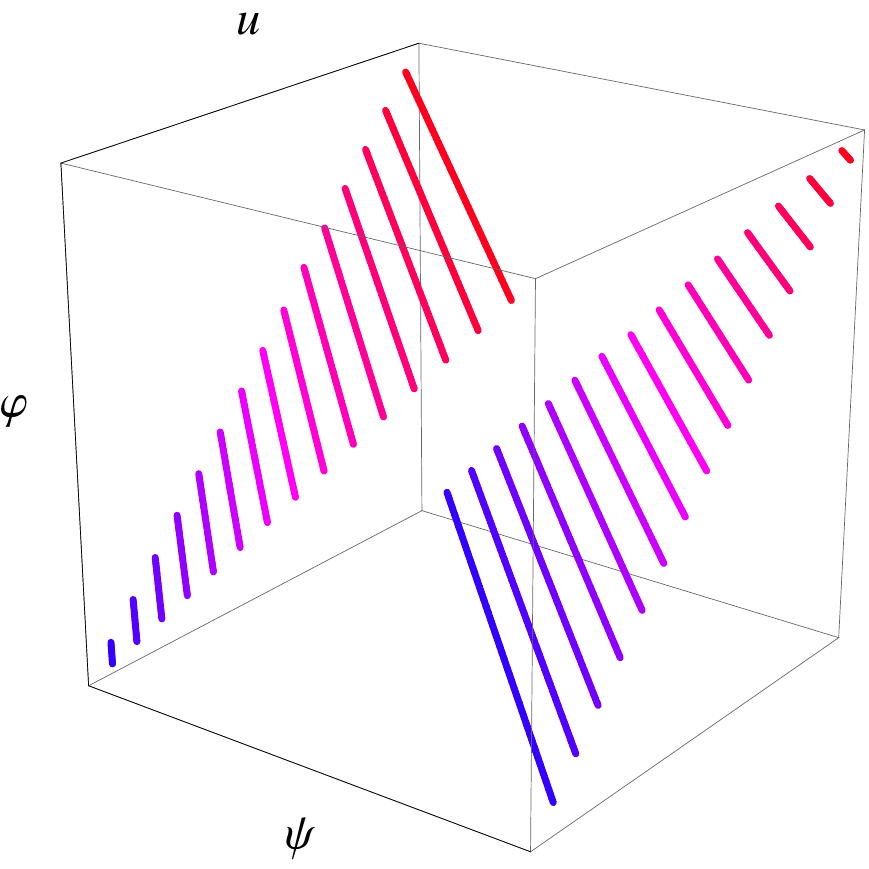} ~ ~
~ ~ \includegraphics[width=45mm]{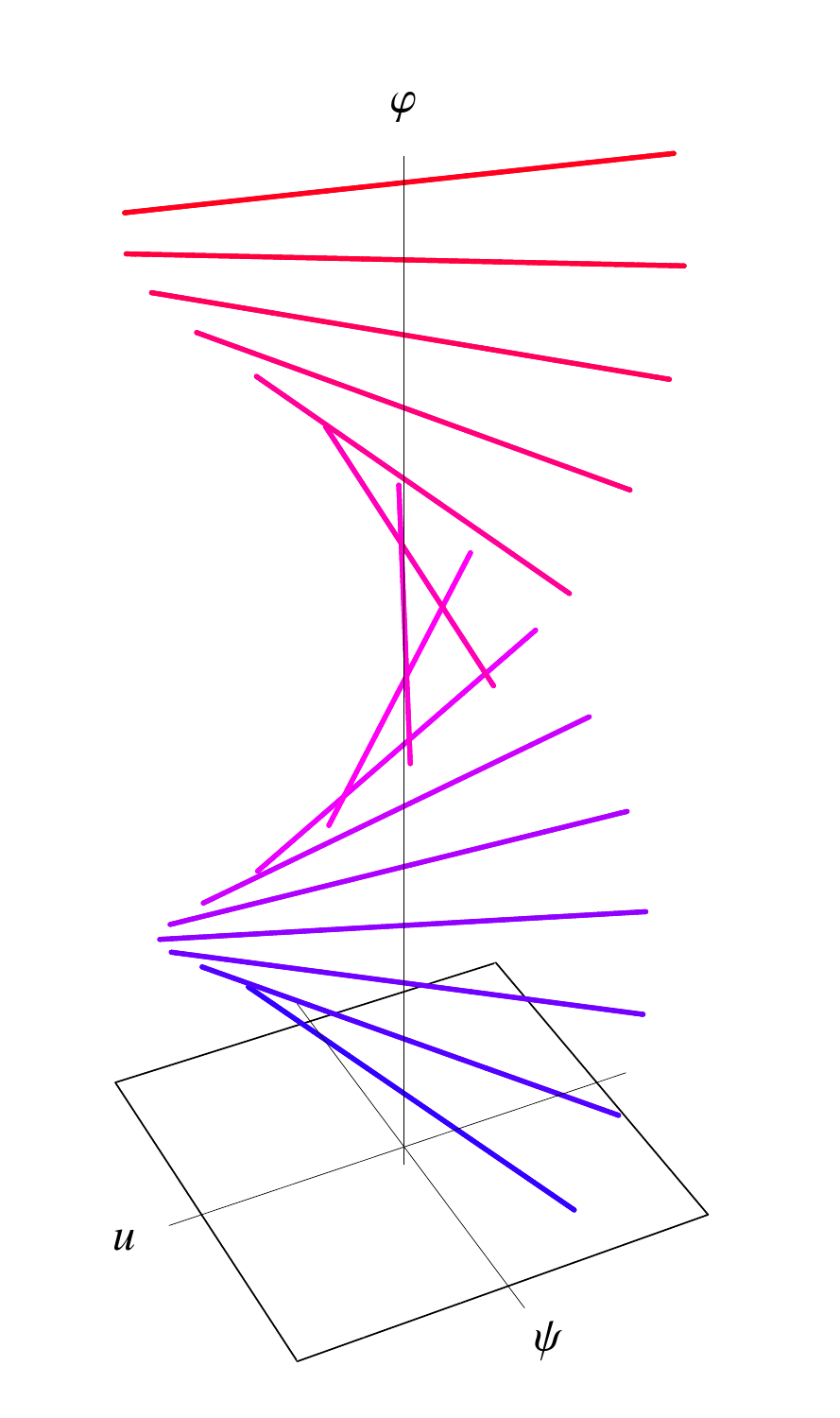} ~ ~ ~ ~ \includegraphics[width=35mm]{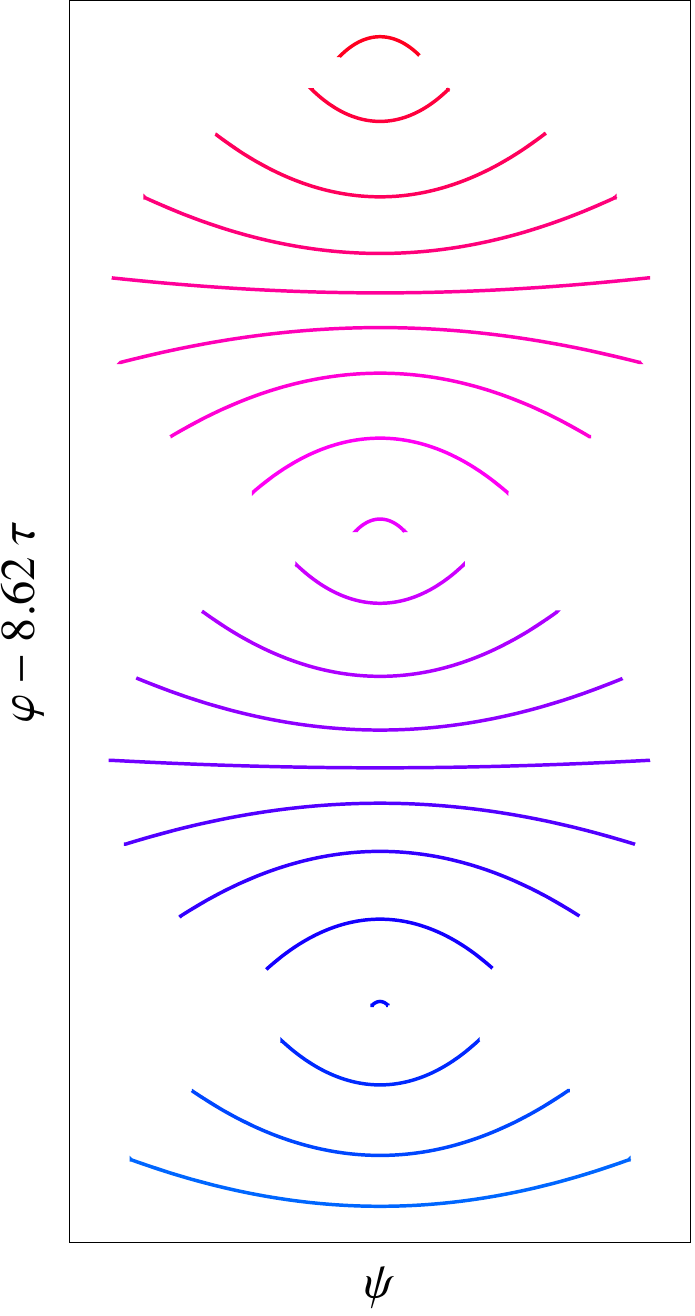} 

\protect\caption{The circular string \eqref{eq:circular-class-WS}, folded string \eqref{eq:folded-class-WS},
and LS's solution \eqref{eq:class-LS}. In all of these the lines
from blue to red (and up the page) represent increasing time. The
first has zero winding along $\varphi$ (i.e. $m_{\varphi}=0$) and
$m_{\psi}=m_{u}=1$. In the last, we take $\nu=\tilde{\nu}=2$ and
have subtracted off 90\% of the motion along $\varphi$.\label{fig:classical-solns-2-3-5}}
\end{figure}

\item The final example is a folded spinning string, with the centre of
mass moving in the $\varphi$ direction:
\begin{equation}
\psi+i\, u=A\, e^{i\nu\tau}\cos\nu\sigma,\qquad\varphi=\omega\tau,\qquad t=\kappa\tau.\label{eq:folded-class-WS}
\end{equation}
Here $\nu\in\mathbb{Z}$ counts the number of windings. When $A\to0$
this becomes the supersymmetric BMN particle, and when $A\to\infty$
it stops moving along the equator. It has charges 
\begin{equation}
\Delta=4\pi g\:\kappa,\qquad J_{+}=J_{-}=J'=4\pi g\:\omega,\qquad S=2\pi g\: A^{2}\nu\label{eq:folded-charges-WS}
\end{equation}
where $S$ is the angular momentum in the $\psi$-$u$ plane:
\[
S=R^{2}\int\frac{d\sigma}{2\pi}\im\left[(\psi-i\, u)\:\partial_{\tau}(\psi+i\, u)\vphantom{1^{1^{1}}}\right].
\]
The contributions to the Virasoro constraints from particular spheres
are quite complicated, for example:
\begin{align*}
V_{S+}^{\text{diag}} & =\cos^{2}\phi\:\omega^{2}+\sin\phi\cos\phi\: A\omega\nu\:\cos\nu\sigma\:\sin\nu\tau+\sin^{2}\phi\:\frac{A^{2}\nu^{2}}{2}\left(1-\cos2\nu\sigma\:\sin2\nu\tau\right)\\
V_{u}^{\text{diag}} & =\frac{A^{2}\nu^{2}}{2}\left(1+\cos2\nu\sigma\:\sin2\nu\tau\right)
\end{align*}
and off-diagonal
\[
\begin{aligned}V_{S+}^{\text{off}} & =\sin\phi\cos\phi\: A\omega\nu\:\sin\nu\sigma\:\cos\nu\tau+\sin^{2}\phi\:\frac{A^{2}\nu^{2}}{4}\sin2\nu\sigma\:\sin2\nu\tau\\
V_{u}^{\text{off}} & =-\frac{A^{2}\nu^{2}}{4}\sin2\nu\sigma\:\sin2\nu\tau.
\end{aligned}
\]
However the total Virasoro constraints simply impose that the cusps
are lightlike, $\kappa^{2}=\omega^{2}+A^{2}\nu^{2}$, or in terms
of the angular momenta,%
\footnote{The equivalent relation for short spinning strings in AdS has instead
an expansion in $\mathcal{S}$ on the right hand side, $\mathcal{S}+\sum_{n\geq2}\: f_{n}(\mathcal{J})\:\mathcal{S}^{n}$
\cite{Gromov:2011bz}. %
} 
\begin{equation}
\kappa^{2}-\cJ^{2}=\frac{\nu}{2}\cS\label{eq:folded-WS-charge-rel}
\end{equation}
where $\cS=S/\sqrt{\lambda}$. 
\end{enumerate}
One of the solutions studied by \cite{Lloyd:2013wza} is related to
this folded string. Both move in the $\varphi$ direction, but while
our solution rotates in the $\psi$-$u$ plane, theirs is confined
to $u=0$ and thus oscillates in $\varphi$. If we start from something
very similar to \eqref{eq:folded-class-WS}'s $\psi$ (agreeing exactly
when $a=\tilde{a}=A/2$ and $\nu=\tilde{\nu}$): 
\[
\psi=a\,\cos\nu(\tau+\sigma)+\tilde{a}\,\cos\tilde{\nu}(\tau-\sigma)
\]
 and demand $u=0$, then we are led to%
\footnote{Here $E(z\vert m)=\int_{0}^{z}d\theta\sqrt{1-m\sin^{2}\theta}={\tt EllipticE[z,m]}$.
Often $m=k^{2}$. %
} 
\begin{equation}
\varphi=\frac{\kappa}{2\nu}E\Big(\nu(\tau+\sigma)\Big\vert\frac{4a^{2}\nu^{2}}{\kappa^{2}}\Big)+\frac{\kappa}{2\tilde{\nu}}E\Big(\tilde{\nu}(\tau-\sigma)\Big\vert\frac{4\tilde{a}^{2}\tilde{\nu}^{2}}{\kappa^{2}}\Big)\label{eq:class-LS}
\end{equation}
very much like equation (4.48) of \cite{Lloyd:2013wza}.%
\footnote{The co-ordinates used by \cite{Lloyd:2013wza} have a parameter $R$
in order to take the Penrose limit, but this is not necessary for
the solution of section 4.3.2. Setting $R=1$ (everywhere except the
$\smash{R^{2}=\sqrt{\lambda}}$ in front of the action) and $\zeta=\phi$
aligns our co-ordinates perfectly with theirs, up to re-naming $\varphi_{+,-}=\psi_{1,2}$,
$\varphi=\eta$, $\psi=x_{1}$, $u=\chi$. After this \eqref{eq:class-LS}
is exactly their (4.48). %
} For this to be real we need $\kappa>2a\nu,2\tilde{a}\tilde{\nu}$,
and for a closed string we need $a\nu=\tilde{a}\tilde{\nu}$. Plotting
$\varphi$, it has some small fluctuations on top of rapid motion
with $\tau$, as drawn in figure \ref{fig:classical-solns-2-3-5}.
Like our folded string, the contributions to the Virasoro constraint
from each sphere are not constants. For example when $a=\tilde{a}=A/2$,
$\nu=\tilde{\nu}=1$ and $\phi=\tfrac{\pi}{4}$: 
\begin{align*}
V_{S+}^{\text{diag}} & =\frac{\kappa^{2}}{2}+\frac{A}{2}\sin(\sigma+\tau)\sqrt{\kappa^{2}-A^{2}\sin^{2}(\sigma+\tau)}-\frac{A}{2}\sin(\sigma-\tau)\sqrt{\kappa^{2}-A^{2}\sin^{2}(\sigma-\tau)}\\
V_{S+}^{\text{off}} & =\frac{A}{4\sqrt{2}}\sin(\sigma+\tau)\sqrt{\kappa^{2}-A^{2}\sin^{2}(\sigma+\tau)}+\frac{A}{4\sqrt{2}}\sin(\sigma-\tau)\sqrt{\kappa^{2}-A^{2}\sin^{2}(\sigma-\tau)}
\end{align*}
but clearly always $V_{u}=0$. Our solution \eqref{eq:folded-class-WS},
which explores also the $S^{1}$ direction, has the virtue of being
much simpler, because it is rigidly rotating, and this allows us to
calculate its mode frequencies below. We now turn to this problem.

\subsection{Modes of the Circular String}

The bosonic modes of \eqref{eq:circular-class-WS} are easy to find,
since this is a homogenous solution, i.e. $\partial_{\sigma}$ generates
an isometry of target space. We make the following ansatz: \newcommand{\qlf}{\tfrac{1}{\lambda^{1/4}}}
\begin{align}
t & =\kappa\tau+\qlf\tilde{t} & \rho & =0+\qlf\tilde{\rho}\nonumber \\
\varphi_{+} & =\omega_{+}\tau+m_{+}\sigma+\qlf\cos\phi\:\tilde{\varphi}_{+} & \theta_{+} & =\tfrac{\pi}{2}+\qlf\cos\phi\:\tilde{\theta}_{+}\displaybreak[0]\label{eq:WS-pert-ansatz-circular}\\
\varphi_{-} & =\omega_{-}\tau+m_{-}\sigma+\qlf\sin\phi\:\tilde{\varphi}_{-} & \theta_{-} & =\tfrac{\pi}{2}+\qlf\sin\phi\:\tilde{\theta}_{-}\nonumber \\
u & =\omega_{u}\tau+m_{u}\sigma+\qlf\tilde{u}.\nonumber 
\end{align}
We set $\gamma=\beta_{\pm}=0$ on the understanding that $\tilde{\rho}$,
$\tilde{\theta}_{+}$ and $\tilde{\theta}_{-}$ each represent one
of two equivalent directions away from $\rho=0$ and $\theta_{\pm}=\frac{\pi}{2}$.
We have also scaled $\tilde{\varphi}_{\pm},\tilde{\theta}_{\pm}$
so as to produce the canonical kinetic term in the quadratic Lagrangian:
\[
2\mathcal{L}_{2B}=-\partial_{\mu}\tilde{t}\partial^{\mu}\tilde{t}+\sum_{x=\rho,\varphi_{\pm},\theta_{\pm},u}\partial_{\mu}\tilde{x}\partial^{\mu}\tilde{x}\;+\kappa^{2}\tilde{\rho}^{2}+(\omega_{+}^{2}-m_{+}^{2})\tilde{\theta}_{+}^{2}+(\omega_{-}^{2}-m_{-}^{2})\tilde{\theta}_{-}^{2}.
\]
The solutions are all plane waves $\tilde{x}=C_{x}\exp(iw\tau+in\sigma)$
with $w=\sqrt{n^{2}+s^{2}}$, and we read off the following masses:
\begin{equation}
s^{2}=\begin{cases}
\kappa^{2} & 2\: AdS\mbox{ modes }(\tilde{\rho}\mbox{ etc.})\\
\omega_{\pm}^{2}-m_{\pm}^{2} & 4\,\perp\mbox{ sphere modes }(\tilde{\theta}_{\pm}\mbox{ etc.})\\
0 & 4\:\mbox{massless }(\tilde{t},\tilde{\varphi}_{\pm},\tilde{u})\mbox{, of which 2 are gauge.}
\end{cases}\label{eq:WS-masses-bosons-circular}
\end{equation}

Only the massless modes here influence the Virasoro constraints at
leading order, and writing $E=i\, e^{iw_{n}\tau+in\sigma}/\lambda^{1/4}$
(with $w_{n}=\left|n\right|$) the changes are%
\footnote{Latin $w_{n}$ is the mode frequency (with respect to $\tau$ not
$AdS$ time $t$), Greek $\omega_{\pm}$ are the classical angular
momenta. As usual $\omega_{M}^{AB}$ below is the spin connection,
for which $M,N$ are curved and $A,B$ flat target space indices.
And in sections \ref{sec:Algebraic-Curve} and \ref{sec:Energy-Corrections},
$\omega_{n}=\Omega(y_{n})$ is the physical frequency (with respect
to $t$).%
} 
\begin{align}
\delta V_{AdS}^{\text{diag}} & =-2C_{t}\,\kappa\, w_{n}\, E, & \delta V_{AdS}^{\text{off}} & =-C_{t}\,\kappa\, n\, E\nonumber \\
\delta V_{S+}^{\text{diag}} & =2C_{\varphi+}(\omega_{+}w_{n}+m_{+}n)E/\cos\phi, & \delta V_{S+}^{\text{off}} & =C_{\varphi+}(\omega_{+}n+m_{+}w_{n})E/\cos\phi\displaybreak[0]\label{eq:WS-circular-delta-vir}\\
\delta V_{S-}^{\text{diag}} & =2C_{\varphi-}(\omega_{-}w_{n}+m_{-}n)E/\sin\phi, & \delta V_{S-}^{\text{off}} & =C_{\varphi-}(\omega_{-}n+m_{-}w_{m})E/\sin\phi\nonumber \\
\delta V_{u}^{\text{diag}} & =2C_{u}(\omega_{u}w_{n}+m_{u}n)E, & \delta V_{u}^{\text{off}} & =C_{u}(\omega_{u}n+m_{u}w_{n})E.\nonumber 
\end{align}
To preserve the total Virasoro constraints \eqref{eq:virasoro-terms-WS}
we can always solve for (say) $C_{t}$ and $C_{u}$, leaving $\tilde{\varphi}_{\pm}$
as physical modes. Doing so will not make individual contributions
(such as $\smash{\delta V_{S+}^{\text{diag}}}$) constant, but let
us observe that they will integrate to zero: $\smash{\int_{0}^{2\pi}d\sigma\:\delta V_{S+}^{\text{diag}}}=0$
etc. 

The fermionic quadratic Lagrangian is given by \newcommand{\tempepsilon}{$\epsilon^{01}=1$}
\begin{align*}
\hfill\mathcal{L}_{2F} & =\left(\eta^{\mu\nu}\delta^{IJ}-\epsilon^{\mu\nu}\sigma_{3}^{IJ}\right)\bar{\Theta}^{I}\rho_{\mu}D_{\nu}\Theta^{J}\\
\shortintertext{where\ \tempepsilon\ and}\qquad D_{\mu}\Theta^{I} & =\left(\partial_{\mu}+\tfrac{1}{4}\partial_{\mu}X^{M}\omega_{M}^{AB}\Gamma_{AB}\right)\delta^{IJ}\Theta^{J}+F\rho_{\mu}\sigma_{1}^{IJ}\Theta^{J}\displaybreak[0]\\
\rho_{\mu} & =\partial_{\mu}X^{M}E_{M}^{A}\Gamma_{A}\\
F & =\tfrac{1}{24}F_{ABC}^{(3)}\Gamma^{ABC}=\tfrac{6}{24}\left(\Gamma^{012}+\cos\phi\,\Gamma^{345}+\sin\phi\,\Gamma^{678}\right).
\end{align*}
The equations of motion are \cite{Minahan:2007gf,Abbott:2008yp} 
\begin{equation}
(\rho_{0}-\rho_{1})(D_{0}+D_{1})\Theta^{1}=0,\qquad(\rho_{0}+\rho_{1})(D_{0}-D_{1})\Theta^{2}=0.\label{eq:WS-fermi-eqm}
\end{equation}
For the circular string this is all quite simple, since the spin connection
term is zero, and the projections of gamma-matrices have constant
coefficients:
\[
\rho_{0}=\kappa\Gamma_{0}+\frac{\omega_{+}}{\cos\phi}\Gamma_{4}+\frac{\omega_{-}}{\sin\phi}\Gamma_{7}+\omega_{u}\Gamma_{9},\qquad\rho_{1}=\frac{m_{+}}{\cos\phi}\Gamma_{4}+\frac{m_{-}}{\sin\phi}\Gamma_{7}+m_{u}\Gamma_{9}\,.
\]
Making the ansatz $\Theta^{I}=e^{i(w_{n}\tau+n\sigma)}\Theta_{0}^{I}$
and fixing $\kappa$-symmetry by defining $\Psi_{0}=(\rho_{0}-\rho_{1})\Theta_{0}^{1}$,
we get  
\begin{equation}
(w_{n}^{2}-n^{2})\Psi_{0}=(\rho_{0}-\rho_{1})F(\rho_{0}+\rho_{1})F\Psi_{0}\,.\label{eq:WS-fermi-eigenvalue-rhoRrhoR}
\end{equation}
The eigenvalues of the matrix on the right hand side give the following
masses: 
\begin{equation}
s^{2}=\frac{1}{4}\begin{cases}
(\kappa-\omega_{+}-\omega_{-})^{2}-(m_{+}+m_{-})^{2}\\
(\kappa+\omega_{+}-\omega_{-})^{2}-(m_{+}-m_{-})^{2}\\
(\kappa-\omega_{+}+\omega_{-})^{2}-(m_{+}-m_{-})^{2}\\
(\kappa+\omega_{+}+\omega_{-})^{2}-(m_{+}+m_{-})^{2}.
\end{cases}\label{eq:WS-masses-fermions-circular}
\end{equation}

Note that both bosonic and fermionic masses are independent of $\phi$,
and free of $\omega_{u}$ and $m_{u}$. Thus the only effect of momentum
$\omega_{u}$ in the $S^{1}$ is through the Virasoro constraints,
i.e. in allowing different choices for $\omega_{\pm}$ compared to
the $\omega_{u}=0$ case.

\subsection{Modes of the Point Particle}

\begin{figure}
\centering \includegraphics[width=60mm]{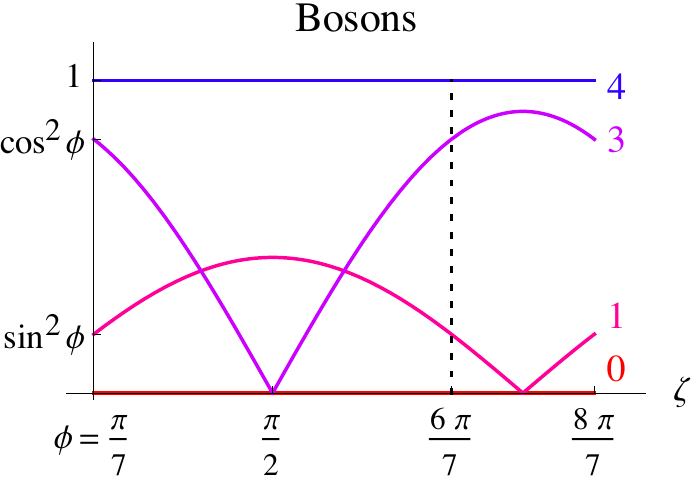} ~ ~ ~ ~
\includegraphics[width=60mm]{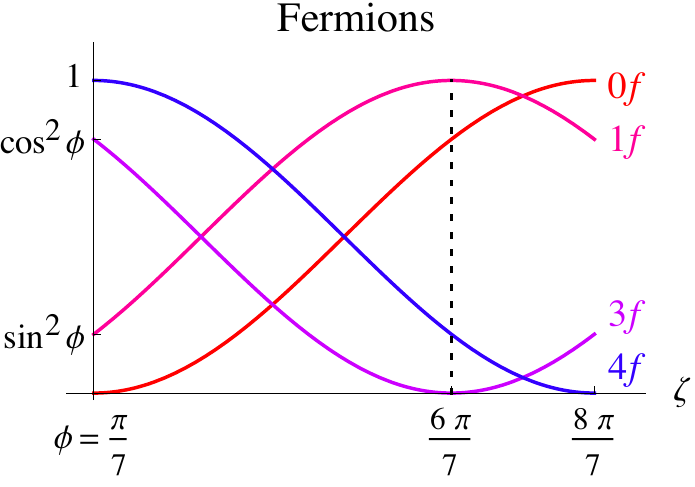} 

\protect\caption{Masses $s/\kappa$ from \eqref{eq:WS-masses-point} of modes of point
particle solutions \eqref{eq:vacuum-zeta-class-WS}, for $\phi=\pi/7$.
The two extremes in the range of $\zeta$ drawn are BMN and a similar
solution moving in the opposite direction, both of which are supersymmetric.
The point of this is to illustrate that the fermions are re-organised
by rotating $\zeta$ from one to the other. Notice that there is a
third supersymmetric solution at $\zeta=\pi-\phi$ (and similarly
a fourth one at $\zeta=-\phi$, not shown). \label{fig:modes-of-point-particle-zeta}}
\end{figure}
This solution \eqref{eq:vacuum-zeta-class-WS} is just the special
case $m_{\pm}=m_{u}=0$ and $\omega_{u}=0$ of the circular string
\eqref{eq:circular-class-WS} treated above. Writing the masses \eqref{eq:WS-masses-bosons-circular}
and \eqref{eq:WS-masses-fermions-circular} in terms of angles $\phi$
and $\zeta$, and labelling them as in the algebraic curve, we have%
\footnote{These fermionic masses were also calculated by \cite{Forini:2012bb}.%
} 
\begin{align}
s_{0} & =0 & s_{0f} & =\tfrac{\kappa}{2}(1-\cos\phi\cos\zeta-\sin\phi\sin\zeta)\nonumber \\
s_{1} & =\kappa\left|\sin\phi\sin\zeta\right| & s_{1f} & =\tfrac{\kappa}{2}(1-\cos\phi\cos\zeta+\sin\phi\sin\zeta)\displaybreak[0]\label{eq:WS-masses-point}\\
s_{3} & =\kappa\left|\cos\phi\cos\zeta\right| & s_{3f} & =\tfrac{\kappa}{2}(1+\cos\phi\cos\zeta-\sin\phi\sin\zeta)\nonumber \\
s_{4} & =\kappa & s_{4f} & =\tfrac{\kappa}{2}(1+\cos\phi\cos\zeta+\sin\phi\sin\zeta).\nonumber 
\end{align}
The supersymmetric BMN case is $\zeta=\phi$. Notice that as we rotate
the direction in which the particle travels, the fermionic modes re-organise
(see figure \ref{fig:modes-of-point-particle-zeta}). Increasing $\zeta$
to $\phi+\pi$, the heaviest fermion becomes massless (and vice versa),
and we again have a supersymmetric solution. What we have done is
to reverse the direction of motion of the string on the sphere: notice
that the charges $J_{\pm}$ in \eqref{eq:vacuum-charges-WS} are both
reversed. 

While we will focus on the cases $\zeta=\phi$ (BMN) and $\zeta=\phi+\pi$
(reversed BMN), note that there are in all four supersymmetric points,
as we recover the same list of masses at $\zeta=-\phi$ and $\zeta=\pi-\phi$.
Physically these solutions reverse the direction of motion on just
one $S^{3}$ (compared to BMN), visible in \eqref{eq:vacuum-charges-WS}.
In these cases the heavy and massless fermions of BMN become light
modes (and vice versa).

\subsection{Bosonic Modes of the Folded String\label{sub:Bosonic-Modes-of-Folded-String}}

To treat the string \eqref{eq:folded-class-WS} instead, it is simplest
to work in Cartesian co-ordinates for the plane in which it rotates,
i.e. to use $\psi$,$u$. Compared to the ansatz for the circular
case \eqref{eq:WS-pert-ansatz-circular} above, we need to replace
$\varphi_{\pm},u$ with 
\begin{align*}
\varphi & =\omega\tau+\qlf\tilde{\varphi}\\
\psi & =A\,\cos\nu\tau\:\cos\nu\sigma+\qlf\tilde{\psi}\\
u & =A\,\sin\nu\tau\:\cos\nu\sigma+\qlf\tilde{u}
\end{align*}
and this leads to 
\begin{equation}
2\mathcal{L}_{2B}=-\partial_{\mu}\tilde{t}\partial^{\mu}\tilde{t}+\sum_{x=\rho,\theta_{\pm},\varphi,\psi,u}\partial_{\mu}\tilde{x}\partial^{\mu}\tilde{x}\;+\kappa^{2}\tilde{\rho}^{2}+s_{\theta+}^{2}\tilde{\theta}_{+}^{2}+s_{\theta-}^{2}\tilde{\theta}_{-}^{2}.\label{eq:L2-folded}
\end{equation}
As expected $\tilde{\psi},\tilde{u}$ remain massless, and the $AdS$
mode $\tilde{\rho}$ has mass $\kappa$.  However the mass terms
for the sphere modes $\tilde{\theta}_{\pm}$ are quite complicated.
This is not unexpected, as the classical solution breaks translational
invariance on the worldsheet. Written in terms of $\sigma^{\pm}=\tfrac{1}{2}(\sigma\pm\tau)$,
we notice that they factorise: 
\begin{align*}
s_{\theta+}^{2} & =\Big(\omega\,\cos^{2}\phi+A\nu\,\sin\phi\cos\phi\,\sin2\nu\sigma^{+}\Big)\Big(\omega\,\cos^{2}\phi-A\nu\,\sin\phi\cos\phi\,\sin2\nu\sigma^{-}\Big)\\
s_{\theta-}^{2} & =\Big(\omega\,\sin^{2}\phi-A\nu\,\sin\phi\cos\phi\,\sin2\nu\sigma^{+}\Big)\Big(\omega\,\sin^{2}\phi+A\nu\,\sin\phi\cos\phi\,\sin2\nu\sigma^{-}\Big).
\end{align*}
This points the way to solving the equations of motion $\partial_{+}\partial_{-}\tilde{\theta}=s_{\theta}^{2}\tilde{\theta}$
by using a separation of variables ansatz $\tilde{\theta}=P(\sigma^{+})M(\sigma^{-})$.
We find the following solution, with $K$ the separation constant:
\begin{align}
\tilde{\theta}_{+}(\sigma,\tau) & =C\, e^{\frac{\omega\,\cos^{2}\phi}{2}\left(K+\frac{1}{K}\right)\sigma+\frac{\omega\,\cos^{2}\phi}{2}\left(K-\frac{1}{K}\right)\tau-A\frac{\sin\phi\cos\phi}{2}\left(K\cos2\nu\sigma^{+}-\frac{1}{K}\cos2\nu\sigma^{-}\right)}\displaybreak[0]\nonumber \\
 & =C\exp\left[in\sigma+iw_{n}\tau+\frac{iA\tan\phi}{\omega}\left(n\sin\nu\sigma\sin\nu\tau-w_{n}\cos\nu\sigma\cos\nu\tau\right)\right]\label{eq:folded-theta-pm}\\
 & =C\exp\Big(i\, n\,\Sigma_{\frac{A}{\omega}\tan\phi}+i\, w_{n}\, T_{\frac{A}{\omega}\tan\phi}\Big),\qquad w_{n}=\sqrt{n^{2}+\omega^{2}\cos^{4}\phi}.\nonumber 
\end{align}
On the second line we solve $\frac{\omega\,\cos^{2}\phi}{2}\left(K+\frac{1}{K}\right)=i\, n$
with $n\in\mathbb{Z}$ to make it periodic in $\sigma$, and on the
third we define 
\begin{equation}
\Sigma_{A'}=\sigma+A'\:\sin\nu\sigma\sin\nu\tau,\qquad T_{A'}=\tau-A'\:\cos\nu\sigma\cos\nu\tau.\label{eq:folded-Sigma-T}
\end{equation}
The modes on the other sphere are similar: 
\[
\tilde{\theta}_{-}(\sigma,\tau)=C\exp\Big(i\, n\,\Sigma_{-\frac{A}{\omega}\cot\phi}+i\, w_{n}\, T_{-\frac{A}{\omega}\cot\phi}\Big),\qquad w_{n}=\sqrt{n^{2}+\omega^{2}\sin^{4}\phi}.
\]

We view the solution \eqref{eq:folded-theta-pm} as being a plane
wave of mass $\omega\cos^{2}\phi$ in an unusual gauge \eqref{eq:folded-Sigma-T}.
Performing the change of variables $(\sigma,\tau)\to(\Sigma_{A\tan\phi/\omega},T_{A\tan\phi/\omega})$
in the action produces the expected mass term:
\begin{equation}
\int d\tau\, d\sigma\left[\partial_{\mu}\tilde{\theta}_{+}\partial^{\mu}\tilde{\theta}_{+}+s_{\theta+}^{2}\tilde{\theta}_{+}^{2}\right]=\int dT\, d\Sigma\left[-\partial_{T}\tilde{\theta}_{+}\partial_{T}\tilde{\theta}_{+}+\partial_{\Sigma}\tilde{\theta}_{+}\partial_{\Sigma}\tilde{\theta}_{+}+\omega^{2}\cos^{4}\phi\:\tilde{\theta}_{+}^{2}\right]\label{eq:folded-action-Sigma-Tee}
\end{equation}
since we have (writing $\Sigma^{M}$ for the new co-ordinates, $M=0,1$)
\[
\det\left(\frac{\partial}{\partial\sigma^{\mu}}\Sigma_{\frac{A}{\omega}\tan\phi}^{M}\right)=\frac{s_{\theta+}^{2}}{\omega^{2}\cos^{4}\phi}.
\]
In these new worldsheet co-ordinates the modes are orthogonal, which
they are not in terms of the original $(\sigma,\tau)$: 
\[
\int_{0}^{2\pi}d\Sigma\re\left[\tilde{\theta}_{+}(n)\right]\re\left[\tilde{\theta}_{+}(n')\right]\propto\delta_{n\, n'}.
\]
The appropriate Hamiltonian is then the one in terms of these variables,
in which time translation is a symmetry of \eqref{eq:folded-action-Sigma-Tee}.
Notice that there is a different $(\Sigma,T)$ for the modes in each
$S^{3}$ (and different again for their superpartners, below).

Working out the effect of modes on the Virasoro constraint as in \eqref{eq:WS-circular-delta-vir},
again only the massless modes matter. The linear variations are 
\begin{align}
\delta V^{\text{diag}} & =-2\kappa\:\partial_{\tau}\tilde{t}+2\omega\:\partial_{\tau}\tilde{\varphi}+2A\nu\left(\cos\nu\sigma\cos\nu\tau\:\partial_{\tau}\tilde{u}-\sin\nu\sigma\sin\nu\tau\:\partial_{\sigma}\tilde{u}\right)\nonumber \\
 & \qquad\qquad\qquad\qquad\quad-2A\nu\left(\cos\nu\sigma\sin\nu\tau\:\partial_{\tau}\tilde{\psi}+\sin\nu\sigma\cos\nu\tau\:\partial_{\sigma}\tilde{\psi}\right)\label{eq:folded-delta-Vir}\\
\delta V^{\text{off}} & =-\kappa\:\partial_{\sigma}\tilde{t}+\omega\:\partial_{\sigma}\tilde{\varphi}-A\nu\left(\sin\nu\sigma\sin\nu\tau\:\partial_{\tau}\tilde{u}-\cos\nu\sigma\cos\nu\tau\:\partial_{\sigma}\tilde{u}\right)\nonumber \\
 & \qquad\qquad\qquad\qquad\;-A\nu\left(\sin\nu\sigma\cos\nu\tau\:\partial_{\tau}\tilde{\psi}+\cos\nu\sigma\sin\nu\tau\:\partial_{\sigma}\tilde{\psi}\right).\nonumber 
\end{align}
These constraints should (as always) kill two modes, leaving two physical
massless modes. All must obey the massless wave equation $\partial_{+}\partial_{-}\tilde{x}=0$,
but they cannot all be plane waves. There are nevertheless solutions
which respect these constraints. For example if $\tilde{\psi}$ is
a massless plane wave with $w_{n}=+n$, its effect can be cancelled
by $\tilde{\varphi}$ as follows: 
\begin{equation}
\tilde{\psi}=C_{\psi}e^{in(\sigma+\tau)},\qquad\tilde{\varphi}(\sigma^{+})=-\frac{C_{\psi}A\nu\: n}{\omega(n^{2}-\nu^{2})}\: e^{in(\sigma+\tau)}\left(n\cos2\nu\sigma^{+}-i\nu\sin2\nu\sigma^{+}\right).\label{eq:folded-massless-psi-strange-phi}
\end{equation}
This $\tilde{\varphi}$ mode is a plane wave multiplied by another
modulating factor, different to the one appearing in \eqref{eq:folded-theta-pm}
but performing a similar role.

\subsection{Fermionic Modes of the Folded String\label{sub:Fermionic-Modes-of-Folded-String}}

The fermions can be analysed in a similar way. The spin connection
term again vanishes, but the $\rho_{\mu}$ are more complicated than
those for the circular string. Let us write them as follows: 
\begin{align*}
\rho_{\pm}=\rho_{0}\pm\rho_{1} & =\kappa\Gamma_{0}+R_{4\pm}(\sigma^{\pm})\:\Gamma_{4}+R_{7\pm}(\sigma^{\pm})\:\Gamma_{7}+R_{9}(\sigma^{\pm})\:\Gamma_{9}\\
\shortintertext{with}R_{4\pm}(\sigma^{\pm}) & =\omega\cos\phi\pm A\nu\sin\phi\,\sin2\nu\sigma^{\pm}\\
R_{7\pm}(\sigma^{\pm}) & =\omega\sin\phi\mp A\nu\cos\phi\,\sin2\nu\sigma^{\pm}\\
R_{9}(\sigma^{\pm}) & =A\nu\cos2\nu\sigma^{\pm}
\end{align*}

The equations of motion \eqref{eq:WS-fermi-eqm} can be written 
\[
\rho_{-}\partial_{+}\Theta^{1}+\rho_{-}F\rho_{+}\Theta^{2}=0,\qquad\rho_{+}\partial_{-}\Theta^{2}+\rho_{+}F\rho_{-}\Theta^{1}=0.
\]
Now multiply from the left with $F$, and define some projected spinors%
\footnote{Both $\rho_{\pm}$ and $F$ are half-rank, and thus this projection
leaves the expected 8 degrees of freedom in $\Phi^{1}$ and $\Phi^{2}$. %
} 
\[
\Phi^{1}=F\rho_{-}\Theta^{1},\qquad\Phi^{2}=F\rho_{+}\Theta^{2}
\]
and also the following functions: 
\[
G_{\pm}(\sigma^{\pm})=F\rho_{\pm}+\rho_{\pm}F=\tfrac{1}{2}\left(-\kappa\Gamma^{12}+\cos\phi\: R_{4\pm}\:\Gamma^{35}+\sin\phi\: R_{7\pm}\:\Gamma^{68}\right).
\]
Then using that $F^{2}=0$ we obtain 
\[
\partial_{+}\Phi^{1}+G_{-}\Phi^{2}=0,\qquad\partial_{-}\Phi^{2}+G_{+}\Phi^{1}=0
\]
and thus $\partial_{+}\partial_{-}\Phi^{I}+G_{+}G_{-}\Phi^{I}=0$.
The matrices $G_{+}$ and $G_{-}$ commute, and we may expand the
solution in their common eigenspinors $\epsilon_{r}$, which are constants:
\[
G_{\pm}(\sigma^{\pm})\:\epsilon_{r}=g_{\pm r}(\sigma^{\pm})\:\epsilon_{r},\qquad\Phi^{I}(\sigma,\tau)=\sum_{r}\tilde{\phi}_{r}^{I}(\sigma,\tau)\:\epsilon_{r}.
\]
Then we obtain a factorised equation for the coefficient fields $\tilde{\phi}_{r}^{I}$
\begin{equation}
\partial_{+}\partial_{-}\tilde{\phi}_{r}^{I}=g_{+r}g_{-r}\tilde{\phi}_{r}^{I}\label{eq:folded-fermi-scalar-eq}
\end{equation}
in terms of the eigenvalues of $G_{\pm}$ 
\[
g_{\pm r}(\sigma^{\pm})=\frac{i}{2}\Big[\alpha_{r}\kappa+\beta_{r}\cos\phi\: R_{4\pm}(\sigma^{\pm})+\gamma_{r}\sin\phi\: R_{7\pm}(\sigma^{\pm})\Big]
\]
in which all eight choices of the signs $\alpha_{r},\beta_{r},\gamma_{r}=\pm1$
occur. 

We can solve \eqref{eq:folded-fermi-scalar-eq} in the same way as
we did the bosonic equation for $\tilde{\theta}_{\pm}$, and for the
cases with signs $\beta_{r}\gamma_{r}=+1$ the result is simply a
plane wave: 
\begin{equation}
\tilde{\phi}_{r}^{I}=C\,\exp\left(in\sigma+iw_{n}\tau\right),\qquad w_{n}=\sqrt{n^{2}+s_{r}^{2}},\qquad s_{r}=\frac{1}{2}\begin{cases}
\kappa-\omega, & r=0f\\
\kappa+\omega, & r=4f\,.
\end{cases}\label{eq:folded-fermi-soln-simple}
\end{equation}
The remaining cases are similar to \eqref{eq:folded-theta-pm}: 
\begin{equation}
\tilde{\phi}_{r}^{I}=C\,\exp\left(in\tilde{\Sigma}_{\frac{A\sin2\phi}{2s_{r}}}+iw_{n}\tilde{T}_{\frac{A\sin2\phi}{2s_{r}}}\right),\qquad s_{r}=\frac{1}{2}\begin{cases}
\kappa-\cos2\phi\:\omega, & r=1f\\
\kappa+\cos2\phi\:\omega, & r=3f
\end{cases}\label{eq:folded-fermi-soln-messy}
\end{equation}
where $\tilde{\Sigma}_{A'}=\sigma+A'\cos\nu\sigma\cos\nu\tau$ and
$\tilde{T}_{A'}=\tau-A'\sin\nu\sigma\sin\nu\tau$. The same comments
made for the bosonic modes about the exotic gauge clearly also apply
here. 

When $A\to0$ these modes all reduce to those of the BMN solution
in \eqref{eq:WS-masses-point}. We have labelled them with names which
make more sense in the algebraic curve formalism below.

\section{Algebraic Curve\label{sec:Algebraic-Curve}}





\newcommand{\pLS}{\smash{\tilde{p}}}
\newcommand{\kLS}{\smash{\tilde{\kappa}}}
\newcommand{\mLS}{\smash{\tilde{m}}}\newcommand{\smalldynkin}[3]{
\begin{tikzpicture}[scale=0.6,semithick]
\draw (1,0) -- (3,0);
\draw [fill=#1] (1,0) circle (2mm); 
\draw [fill=#2] (2,0) circle (2mm); 
\draw [fill=#3] (3,0) circle (2mm); 
\draw [thin] (18mm,0) -- (22mm,0);
\draw [thin] (2,-2mm) -- (2,2mm);
\end{tikzpicture}
} 

\newcommand{\smallcirc}[1]{\begin{tikzpicture}[scale=0.6] \draw [fill=#1] (0,0) circle (2mm); \end{tikzpicture}}

We begin by discussing the recent work of Lloyd and Stefa\'{n}ski
\cite{Lloyd:2013wza}, who study (in section 4) a string moving in
$\mathbb{R}\times S^{1}\times S^{1}$, with metric \eqref{eq:metric-ads3s3s3s1}
at $\theta_{\pm}=\pi/2$, $\rho=0$ and $u=0$. The algebraic curve
for this solution has three quasimomenta, and to avoid a clash of
notation let us call their quasimomenta $\pLS_{l}$, with $l=0,1,2$
describing time, one sphere, and the other sphere. Their residues
are parameterised by $\kLS_{l}$ and $\mLS_{l}$: 
\[
\pLS_{l}(x)=\frac{\kLS_{l}x+2\pi\mLS_{l}}{x^{2}-1}
\]
which are given by 
\begin{align}
\kLS_{0} & =i\int_{0}^{2\pi}d\sigma\:\partial_{\tau}t, & \kLS_{1} & =\frac{-1}{\cos\phi}\int_{0}^{2\pi}d\sigma\:\partial_{\tau}\varphi_{+}, & \kLS_{2} & =\frac{-1}{\sin\phi}\int_{0}^{2\pi}d\sigma\:\partial_{\tau}\varphi_{-}\label{eq:LS-residue-integrals}\\
\mLS_{0} & =0, & 2\pi\mLS_{1} & =\frac{-1}{\cos\phi}\int_{0}^{2\pi}d\sigma\:\partial_{\sigma}\varphi_{+}, & 2\pi\mLS_{2} & =\frac{-1}{\sin\phi}\int_{0}^{2\pi}d\sigma\:\partial_{\sigma}\varphi_{-}\,.\nonumber 
\end{align}
They showed that this is the complete algebraic curve for any solution
in this spacetime, with only poles at $x=\pm1$, never branch cuts.

The traditional Virasoro condition on the residues, applied to this
case, reads 
\begin{equation}
\sum_{l=0}^{D}(\kLS_{l}\pm2\pi\mLS_{l})^{2}=0,\qquad D=2.\label{eq:AC-vir-traditional}
\end{equation}
 The weaker generalised residue condition (GRC) proposed by \cite{Lloyd:2013wza}
is that there exist $f_{l}^{\pm}(\sigma)$ such that 
\begin{equation}
\kLS_{l}\pm2\pi\mLS_{l}=2\int_{0}^{2\pi}d\sigma\: f_{l}^{\pm}(\sigma),\qquad\sum_{l=0}^{D}(f_{l}^{\pm})^{2}=0.\label{eq:AC-vir-new-GRC}
\end{equation}
Clearly if \eqref{eq:AC-vir-traditional} is satisfied then we can
simply take $f$ to be constant. Comments:
\begin{itemize}
\item The new condition is very close to being the worldsheet one. If we
define 
\[
f_{l}^{\pm}=G_{l}(\partial_{\tau}\varphi_{l}\pm\partial_{\sigma}\varphi_{l})
\]
writing $(G_{l},\varphi_{l})=(i,t),\;(\tfrac{1}{\cos\phi},\varphi_{+}),\;(\tfrac{1}{\sin\phi},\varphi_{-})$
for $l=0,1,2$, then the terms of the worldsheet Virasoro constraints
\eqref{eq:virasoro-terms-WS} are 
\begin{align*}
V_{l}^{\text{diag}} & =\frac{(f_{l}^{+})^{2}+(f_{l}^{-})^{2}}{2}=G_{l}^{2}(\partial_{\tau}\varphi_{l})^{2}+G_{l}^{2}(\partial_{\sigma}\varphi_{l})^{2}\\
V_{l}^{\text{off}} & =\frac{(f_{l}^{+})^{2}-(f_{l}^{-})^{2}}{4}=G_{l}^{2}\partial_{\tau}\varphi_{l}\:\partial_{\sigma}\varphi_{l}
\end{align*}
and thus \eqref{eq:virasoro-terms-WS}, i.e. $\sum_{l}V_{l}^{\text{diag}}=0=\sum_{l}V_{l}^{\text{off}}$,
is equivalent to \eqref{eq:AC-vir-new-GRC}. However this $f_{l}^{\pm}$
is a function of $\tau$ as well as $\sigma$. 
\item That the new condition \eqref{eq:AC-vir-new-GRC} does not imply the
traditional one \eqref{eq:AC-vir-traditional} is a consequence of
there being more than two factors in the spacetime. In fact without
saying what $f$ means, the following implication is true for $D=1$
\begin{equation}
0=\sum_{l=0}^{D}f_{l}^{2}(\sigma)\quad\Rightarrow\quad0=\sum_{l=0}^{D}\left[\int d\sigma\: f_{l}(\sigma)\right]^{2}\label{eq:two-vir-toy}
\end{equation}
since $f_{0}=\pm if_{1}$, but fails for $D\geq2$. 
\item Let us also mention here that if we consider a point particle solution
in $\mathbb{R}\times(S^{1})^{D}$, then for $D=1$ there are just
2 solutions (lightlike and moving one way or the other) while in $D\geq2$
there is a continuous set of them, allowing us to rotate from one
direction to the reverse, as in figure \ref{fig:modes-of-point-particle-zeta}. 
\end{itemize}

\subsection{Ten Dimensions}

All string solutions in $\mathbb{R}\times S^{1}\times S^{1}$ will
also be solutions in $AdS_{3}\times S^{3}\times S^{3}$, and thus
we can map their algebraic curves into the one describing the full
space. (Our notation here follows \cite{Abbott:2012dd} closely.)
This has six quasimomenta $p_{\ell}(x)$ with $\ell=1,2,3,\bar{1},\bar{2},\bar{3}$,
corresponding to the six Cartan generators of $d(2,1;\alpha)$. The
map is as follows:
\begin{equation}
\begin{pmatrix}p_{1}\\
p_{2}\\
p_{3}
\end{pmatrix}=\begin{pmatrix}\frac{i}{2}\,\pLS_{0}-\frac{1}{2\sin\phi}\,\pLS_{2}\\
i\,\pLS_{0}\\
\frac{i}{2}\,\pLS_{0}-\frac{1}{2\cos\phi}\,\pLS_{1}
\end{pmatrix},\qquad p_{\bar{\ell}}(\tfrac{1}{x})=p_{\ell}(x).\label{eq:my-map-p-from-pLS}
\end{equation}
Inversion symmetry is $p_{\ell}(\tfrac{1}{x})=S_{\ell m}p_{m}(x)$
with\newcommand{\tempmatrix}{\big[\begin{smallmatrix}0&1\\ 1&0\end{smallmatrix}\big]}
$S=1_{3\times3}\otimes\tempmatrix$. The Cartan matrix is 
\[
A=\left[\begin{array}{ccc}
4\sin^{2}\phi & -2\sin^{2}\phi & 0\\
-2\sin^{2}\phi & 0 & -2\cos^{2}\phi\\
0 & -2\cos^{2}\phi & 4\cos^{2}\phi
\end{array}\right]\otimes1_{2\times2}.
\]
for which we draw the Dynkin diagram \smalldynkin{white}{white}{white},
left and right.%
\footnote{Away from the classical limit we should, according to \cite{Borsato:2012ud,Borsato:2012ss},
use a different fermionic grading on the right. But this makes no
difference to the classical algebraic curve, and thus it is simpler
not to do so here. This was also briefly discussed in \cite{Abbott:2013mpa}. %
} If we parameterise the poles at $\pm1$ by $p_{\ell}(x)=(\kappa_{\ell}x+2\pi\, m_{\ell})/(x^{2}-1)$
as before, then the traditional condition on the residues is \cite{Zarembo:2010yz}%
\footnote{In \eqref{eq:AC-vir-traditional}, instead of taking $\kLS_{0}$ to
be imaginary we could insert $A=\diag(-1,1,1)$ as the Cartan matrix. %
} 
\begin{equation}
(\kappa_{\ell}\pm2\pi m_{\ell})A_{\ell n}(\kappa_{n}\pm2\pi m_{n})=0.\label{eq:AC-vir-with-cartan}
\end{equation}

The angular momentum and winding (or worldsheet momentum) are given
by the behaviour of the quasimomenta at infinity. In terms of $J_{\ell}$
defined as 
\[
p_{\ell}(x)\to P_{\ell}+\frac{1}{2gx}J_{\ell}+\mathcal{O}\Big(\frac{1}{x^{2}}\Big).
\]
the combinations of interest are
\begin{align*}
\Delta & =\tfrac{1}{2}(-J_{2}+J_{\bar{2}}), & J_{+} & =(J_{3}-J_{\bar{3}})-\tfrac{1}{2}(J_{2}-J_{\bar{2}})\\
S & =\tfrac{1}{2}(-J_{2}-J_{\bar{2}}), & J_{-} & =(J_{1}-J_{\bar{1}})-\tfrac{1}{2}(J_{2}-J_{\bar{2}})
\end{align*}
and $J'=\cos^{2}\phi\: J_{+}+\sin^{2}\phi\: J_{-}$. The combinations
of $P_{\ell}$ of interest are discussed at \eqref{eq:P-nodes-zeta}
below. 

In \cite{Abbott:2012dd} another basis of quasimomenta $q_{i}(x)$,
$i=1\ldots6$ was defined, which are more closely related to the ones
usually used in $AdS_{5}\times S^{5}$ or $AdS_{4}\times CP^{3}$,
at least when $\phi=\tfrac{\pi}{4}$. These are defined such that
the Cartan matrix becomes trivial: 
\begin{align*}
\left(\begin{array}{c}
q_{1}\\
\hline q_{3}\\
q_{5}
\end{array}\right) & =B_{\text{left}}\begin{pmatrix}p_{1}\\
p_{2}\\
p_{3}
\end{pmatrix},\qquad B_{\text{left}}=\left[\begin{array}{ccc}
0 & -1 & 0\\
\hline 2\sin^{2}\phi & -1 & 2\cos^{2}\phi\\
-\sin2\phi & 0 & \sin2\phi
\end{array}\right]\\
 & \Rightarrow\quad q_{i}(\mathbb{I}_{1,2})_{ij}q'_{j}=p_{\ell}A_{\ell m}p_{m}'
\end{align*}
writing $\mathbb{I}_{1,2}=\text{diag}(-1,1,1)$ to make the signature
of weight space explicit. (After this we will simply write $q\cdot q'$.)
The right-hand quasimomenta $q_{i\,\text{even}}$ are given by the
inversion symmetry $q_{i}(x)=-q_{i-1}(\tfrac{1}{x})$, and we may
extend to\newcommand{\tempmatrixalt}{\big[\begin{smallmatrix}1&0\\ 0&-1\end{smallmatrix}\big]}
$B=B_{\text{left}}\otimes\tempmatrixalt$ to include them. Then $q_{1}$
and $q_{2}$ describe AdS, while $q_{3}\ldots q_{6}$ describe the
spheres. It is also useful to think of $q_{-i}(x)=-q_{i}(x)$ as being
6 more quasimomenta, to distinguish the signs with which cuts connect
them. (These are analogous to $q_{6}\ldots q_{10}$ in the $AdS_{4}\times CP^{3}$
case, at $\phi=\tfrac{\pi}{4}$.)

Finally we will also need one more quasimomentum for the $S^{1}$
circle, which we can treat on the pattern of \eqref{eq:LS-residue-integrals}
above: 
\begin{equation}
\pLS_{3}(x)=\frac{\kLS_{3}x+2\pi\mLS_{3}}{x^{2}-1},\qquad\kLS_{3}=-\int_{0}^{2\pi}d\sigma\:\partial_{\tau}u,\qquad2\pi\mLS_{3}=-\int_{0}^{2\pi}d\sigma\:\partial_{\sigma}u\,.\label{eq:LS-residue-integral-u}
\end{equation}
We could then extend the sums in \eqref{eq:AC-vir-traditional} and
\eqref{eq:AC-vir-new-GRC} to run up to $l=3$. For now however we
make no attempt to include this as part of the $p_{\ell}$ (but see
section \ref{sub:Massless-Bosons?} below).

\subsection{The Generalised Vacuum}

Zarembo \cite{Zarembo:2010yz} gives two solutions to $\vec{\kappa}A\,\vec{\kappa}=0$,
$S\vec{\kappa}=-\vec{\kappa}$ for the case $\phi=\frac{\pi}{4}$,
i.e. to the Virasoro condition (in the absence of winding) and the
inversion symmetry condition: 
\[
\vec{\kappa}=\left(0,1,0\right)\otimes(1,-1)\qquad\mbox{or }\quad\vec{\kappa}=\left(a,\: a^{2}+(1-a)^{2},\:1-a\right)\otimes(1,-1).
\]
The first of these gives the BMN vacuum. However taking into account
that the normalisation of this $\vec{\kappa}$ is arbitrary, the first
is in fact the $a\to\infty$ case of the second, and thus the BMN
vacuum is part of a one-parameter family. Choosing to use $\zeta$
as the parameter, restoring the $\phi$ dependence, and fixing the
overall normalisation, in this subsection we study the algebraic curve%
\footnote{The vector $\vec{\kappa}$ with components $\kappa_{\ell}$ controls
the poles at $x=\pm1$ in $p_{\ell}(x)$, while $\kLS_{l}$ plays
the same role for $\pLS_{l}(x)$.  The scalar $\kappa$ is the same
constant as in the worldsheet solutions, controlling the energy $\kappa=\Delta/4\pi g$.
Latin $-k_{\ell}$ below controls which nodes are excited by the mode. %
} 
\begin{equation}
p_{\ell}(x)=\frac{x}{x^{2}-1}\kappa_{\ell},\qquad\vec{\kappa}=-\frac{2\pi\kappa}{2}\left(1-\frac{\sin\zeta}{\sin\phi},\;2,\;1-\frac{\cos\zeta}{\cos\phi}\right)\otimes(1,-1)\label{eq:delta-vac-p}
\end{equation}
corresponding to the point particle solution \eqref{eq:vacuum-zeta-class-WS}.
It is also trivial to directly integrate \eqref{eq:vacuum-zeta-class-WS}
using \eqref{eq:LS-residue-integrals} to find $\pLS_{l}$; the map
\eqref{eq:my-map-p-from-pLS} is fixed largely by this comparison.
In the basis $q=Bp$ the same solution reads 
\[
\left(\begin{array}{c}
q_{1}(x)\\
\hline q_{3}(x)\\
q_{5}(x)
\end{array}\right)=\left(\begin{array}{c}
-q_{2}(\tfrac{1}{x})\\
\hline -q_{4}(\tfrac{1}{x})\\
-q_{6}(\tfrac{1}{x})
\end{array}\right)=\frac{x}{x^{2}-1}\:2\pi\kappa\left(\begin{array}{c}
1\\
\hline \cos(\phi-\zeta)\\
\sin(\phi-\zeta)
\end{array}\right)
\]
which clearly solves $q\cdot q=0$ (in $\mathbb{R}^{2,4}$) and inversion
symmetry. Again this reduces to the usual BMN solution when $\zeta=\phi$. 

We can now proceed to construct modes using the method of \cite{Gromov:2007aq,Gromov:2008ec}:
we perturb the quasimomenta by adding new poles at $x=y$ and allowing
the residues at $x=\pm1$ to vary, subject to a condition on the behaviour
at infinity. The perturbation of the energy is $\Omega(y)$, the ``off-shell''
frequency. The mode number $n\in\mathbb{Z}$ fixes the allowed points
$y_{n}$, and thus gives the ``on-shell'' frequencies as $\omega_{n}=\Omega(y_{n})$. 

When we perturb the residues at $x=\pm1$, we should do so in the
most general way allowed by the Virasoro constraint. Since \eqref{eq:delta-vac-p}
is the most general solution (without winding) this means allowing
\begin{equation}
\delta\kappa_{\ell}=\frac{\partial\kappa_{\ell}}{\partial\kappa}\delta\kappa+\frac{\partial\kappa_{\ell}}{\partial\zeta}\delta\zeta\,.\label{eq:delta-vac-delta-kappa}
\end{equation}
The second term here is a new feature,%
\footnote{This explains footnote 2 of \cite{Abbott:2012dd}. At $\zeta=\phi$
the second term of \eqref{eq:delta-vac-delta-kappa} reads $\delta\vec{\kappa}=\smash{\tfrac{\Delta}{4g}}\,\delta\zeta\,\left(\cot\phi,0,-\tan\phi\right)\otimes(1,-1)$.
This contributes to $\delta q$ poles at $x=\pm1$ on sheets 5 and
6 because $B\delta\vec{\kappa}=\smash{\tfrac{-\Delta}{4g}}\,\delta\zeta\,\left(0,0\,\vert\,0,0,1,1\right)$,
which were included there without justification. %
} arising because the Virasoro constraint has a one-parameter family
of solutions \eqref{eq:delta-vac-p}, rather than the discrete solutions
seen in $AdS_{5}\times S^{5}$ and $AdS_{4}\times CP^{3}$. Apart
from this there are no changes to what was done in \cite{Abbott:2012dd}
for the $6+6$ modes considered there, and we recover most of the
masses \eqref{eq:WS-masses-point}: bosons $r=1,3,4$, fermions $r=1f,3f,4f$,
and their barred cousins. We refer to modes $1,1f,3,3f$ as light,
since they have BMN masses $0<s<\kappa$, and modes $4,4f$ as heavy,
BMN mass $\kappa$. 

It will be instructive however to work one example out slowly, and
we focus on the heavy fermion $4f$. This mode turns on nodes $1,2,3$,
which we draw as \smalldynkin{\colourone}{\colourone}{\colourone}
and write as $k_{\ell}=-1$ for $\ell=1,2,3$. Thus we must consider%
\footnote{The residue $\alpha(y)$ is from \cite{Beisert:2005bv}. The second
term in square brackets is a twist like that needed for the giant
magnon in \cite{Gromov:2008ie,Abbott:2013mpa}, although in fact it
plays very little role here. We allow the perturbation to carry momentum
$\delta P_{\ell}$. %
} 
\[
\delta p_{\ell}(x)=\delta\kappa_{\ell}\frac{x}{x^{2}-1}-\left[\frac{\alpha(y)}{x-y}+\frac{1}{2}\frac{\alpha(y)}{y}\right]\qquad\mbox{for}\;\ell=1,2,3
\]
where $\alpha(y)=\frac{1}{2g}\frac{y^{2}}{y^{2}-1}$, and the other
sheets are filled in by inversion symmetry $\delta p_{\bar{\ell}}(x)=\delta p_{\ell}(\tfrac{1}{x})$.
We demand that at infinity, 
\begin{align*}
\delta p_{\ell}(x) & =\delta P_{\ell}+\frac{1}{2gx}\left(-1+D_{\ell}\delta\Delta\right)+\bigodiv{x^{2}},\qquad\ell=1,2,3\\
\delta p_{\bar{\ell}}(x) & =\delta P_{\bar{\ell}}+\frac{1}{2gx}\: D_{\bar{\ell}}\delta\Delta+\ldots
\end{align*}
where $D=B^{-1}(1,1\:\vert\,0,0,0,0)=\smash{\left(-\tfrac{1}{2},-1,-\tfrac{1}{2}\right)}\otimes(1,-1)$
encodes the change in energy. Solving, we find off-shell frequency
\begin{equation}
\Omega_{4f}(y)=\delta\Delta=\frac{\left[1+\cos(\phi-\zeta)\right]}{y^{2}-1}\label{eq:delta-vac-Omega-4f}
\end{equation}
and momentum $\delta P_{\bar{1}}=\delta P_{\bar{3}}=n/4g\kappa$.
To put this mode on shell we must solve for $y$ in terms of the mode
number $n$ in 
\begin{align}
2\pi n_{4f} & =-k_{\ell}A_{\ell m}p_{m}(y)=\left(A_{1m}+A_{2m}+A_{3m}\right)p_{m}(y)\nonumber \\
 & =2\sin^{2}\phi\: p_{1}-2\, p_{2}+2\cos^{2}\phi\: p_{3}\nonumber \\
 & =2\pi\kappa\left[1+\cos(\phi-\zeta)\right]\frac{y}{y^{2}-1}\label{eq:delta-vac-2pin-4f}
\end{align}
giving 
\begin{equation}
y_{n}=\frac{\kappa}{2n}\left[1+\cos(\phi-\zeta)\right]\pm\sqrt{1+\frac{\kappa^{2}}{4n^{2}}\left[1+\cos(\phi-\zeta)\right]^{2}}.\label{eq:delta-vac-yn-4f}
\end{equation}
The physical, on-shell frequency is then 
\begin{equation}
\omega_{n}=\Omega_{4f}(y_{n})=-\frac{1}{2}\left[1+\cos(\phi-\zeta)\right]\pm\sqrt{\frac{n^{2}}{\kappa^{2}}+\frac{1}{4}\left[1+\cos(\phi-\zeta)\right]^{2}}\label{eq:delta-vac-freq-4f}
\end{equation}
and we choose always the positive sign here, which selects the pole
in the physical region $\left|y_{n}\right|>1$. The mass $s_{4f}=\tfrac{\kappa}{2}\left[1+\cos(\phi-\zeta)\right]$
matches the worldsheet calculation \eqref{eq:WS-masses-point}. 

This mode is in fact a little simpler in terms of the basis $q$,
where $4f=(1,-3)$, that is, it connects sheets $q_{1}$ and $q_{-3}=-q_{3}$.
For the light modes we must set $\phi=\pi/4$ to have this simple
interpretation (away from this they influence more than two $q_{i}$)
but for the heavy modes we do not need to do so. The equations above
can be written 
\[
\delta q(x)=\frac{2\pi\: x}{x^{2}-1}\left(\begin{array}{c}
\delta\kappa\\
\delta\kappa\\
\hline c_{-}\,\delta\kappa+s_{-}\,\delta\zeta\\
c_{-}\,\delta\kappa+s_{-}\,\delta\zeta\\
s_{-}\,\delta\kappa-c_{-}\,\delta\zeta\\
s_{-}\,\delta\kappa-c_{-}\,\delta\zeta
\end{array}\right)+\left(\begin{array}{c}
\left[\frac{\alpha(y)}{x-y}+\frac{1}{2}\frac{\alpha(y)}{y}\right]\\
0\\
\hline -\left[\frac{\alpha(y)}{x-y}+\frac{1}{2}\frac{\alpha(y)}{y}\right]\\
0\\
0\\
0
\end{array}\right)\to\frac{1}{2gx}\left(\begin{array}{c}
\delta\Delta+1\\
\delta\Delta\\
\hline -1\\
0\\
0\\
0
\end{array}\right)+\bigodiv{x^{2}}
\]
(writing $c_{-}=\cos(\phi-\zeta)$ and $s_{-}=\sin(\phi-\zeta)$)
and 
\[
2\pi n_{4f}=q_{1}(y)-q_{-3}(y)=q_{1}(y)+q_{3}(y)=2\pi\kappa\left[1+\cos(\phi-\zeta)\right]\frac{y}{y^{2}-1}.
\]
These give exactly the same results \eqref{eq:delta-vac-yn-4f}, \eqref{eq:delta-vac-freq-4f}.

\subsection{Construction of Missing Fermions\label{sub:Construction-of-Missing-Fermions}}

Now recall our discussion of the modes of the point particle in the
worldsheet theory, \eqref{eq:WS-masses-point}. We saw that moving
from $\zeta=\phi$ to $\zeta=\phi+\pi$ (which reverses the direction
of the BMN particle) made the heavy fermion $4f=(1,-3)$ become massless,
and of course we recover this fact here, in \eqref{eq:delta-vac-freq-4f}.
But we also saw that the massless fermion became heavy, and using
this observation we can learn how to describe this mode (which we
shall call ``$0f$'') in the algebraic curve: it must be the mode
which, near to $\zeta=\phi+\pi$, behaves exactly like $4f$ did near
to $\zeta=\phi$. 

In terms of the $q_{i}$ this is fairly obvious: increasing $\zeta\to\zeta+\pi$
changes $q_{3}\to-q_{3}$ and $q_{5}\to-q_{5}$, and thus we want
$0f=(1,3)$. Translating back to the $p_{\ell}$ we find that $0f=\smalldynkin{white}{\colourone}{white}$,
i.e. we turn on only the node $2$. We can then add this to the list
of modes whose frequencies we can calculate by the procedure above.
And it has exactly the mass expected from the worldsheet calculation. 

Let us be a bit more careful: so far we have discussed the effect
of reversing one particular solution (the vacuum) on one mode ($4f$).
We would like to argue that the same idea holds more generally. Consider
now some arbitrary algebraic curve $q(x)=B\, p(x)$, and define from
this a reversed curve\vspace{-2mm} 
\begin{equation}
q_{i}'(x)=\begin{cases}
q_{i}(x), & i=1,2\\
-q_{i}(x), & i=3,4,5,6
\end{cases}\qquad\Leftrightarrow\quad p'=B^{-1}q'=\begin{pmatrix}-p_{1}+p_{2}\\
p_{2}\\
-p_{3}+p_{2}
\end{pmatrix}.\label{eq:reversal-symm-q-and-p}
\end{equation}
This is also a valid solution. It has the same conserved charges in
AdS but all of its sphere angular momenta are minus those of the initial
solution. Then we ask: to reproduce the effect of a given mode $r$
on $q$, what mode $r'$ must we use on $q'$? 

At $\phi=\tfrac{\pi}{4}$ every mode $r=(i,j)$ connects two sheets
$q_{i},q_{j}$. The mode number is 
\[
2\pi n=q_{i}-q_{j}\quad\to\quad2\pi n'=q'_{i'}-q'_{j'}.
\]
For these to be equal, clearly we need $i'=i$ for $i=1,2$ and $i'=-i$
for $i\geq3$. Then AdS bosons are unchanged, and for sphere bosons
we need only change the sign of the mode number, $n'=-n$, which we
can ignore. But for fermions, which connect an AdS sheet to a sphere
one, the change is $(i,j)\to(i',j')=(i,-j)$, i.e.\newcommand{\myarrow}{\ \ \leftrightarrow\ \ }\newcommand{\myarrowone}{\ \ \to\ \ }
\begin{align}
3f=(1,-5) & \myarrow1f=(1,5) & \bar{3}f=(2,-6) & \myarrow\bar{1}f=(2,6)\label{eq:reversal-symm-on-modes}\\
4f=(1,-3) & \myarrow0f=(1,3) & \bar{4}f=(2,-4) & \myarrow\bar{0}f=(2,4).\nonumber 
\end{align}
While we discussed the mode number, it is clear that the same thing
happens for the placement of the new poles (cuts) into sheets: instead
of connecting $i$ to $j$ we connect $i'$ to $j'$ by exactly the
same rules. And while we have written this down in terms of the $q_{i}$,
clearly \eqref{eq:reversal-symm-q-and-p} shows that the rule in terms
of the $p_{\ell}$ is  
\[
\left(k{}_{1},k{}_{2},k{}_{3}\right)\to\left(k'_{1},k'_{2},k'_{3}\right)=\left(-k_{1}+k_{2},\: k_{2},\:-k_{3}+k_{2}\right).
\]
Applied to the list in table \ref{tab:List-of-modes} we obtain the
same map \eqref{eq:reversal-symm-on-modes}, now valid at any $\phi$. 

This map \eqref{eq:reversal-symm-on-modes} takes the six original
fermions (massive for BMN) and gives us a different set of six (massive
for reversed BMN), with some overlap. We argue that the union of these
two sets is precisely the full set of eight physical fermions which
must always exist. For generic classical solutions they will all be
nontrivial, and the simplest example of this is the non-supersymmetric
point particle, \eqref{eq:delta-vac-p}. Perturbing this solution,
the eight fermions $0f\ldots\bar{4}f$ exactly reproduce the worldsheet
masses \eqref{eq:WS-masses-point}. We will see similar agreement
for other classical solutions below. 

\newcommand{\rlapij}{\rlap{$(i,j)$}}

\begin{table}
\centering{}\centering %
\begin{tabular}{lcccc|c}
 & $r$ & BMN mass &  & Constraint on $\delta m_{\ell}$ & At $\phi=\tfrac{\pi}{4}$, $r=\rlapij$\tabularnewline
\hline 
Massless: & $0$ & 0 & --- & --- & ---\tabularnewline
 & $0f$ & 0 & \smalldynkin{white}{\colourone}{white} & $\delta P_{\bar{1}}=\delta P_{\bar{3}}=0$ & $(1,3)$\tabularnewline
\hline 
Light: & $1$ & $\kappa\sin^{2}\phi$ & \smalldynkin{\colourone}{white}{white} & $\delta P_{\bar{2}}=\delta P_{\bar{3}}=0$ & $(3,5)$\tabularnewline
 & $1f$ & $\kappa\sin^{2}\phi$ & \smalldynkin{\colourone}{\colourone}{white} & $\delta P_{\bar{1}}=\delta P_{\bar{2}}$, $\delta P_{\bar{3}}=0$ & $(1,5)$\tabularnewline
 & $3$ & $\kappa\cos^{2}\phi$ & \smalldynkin{white}{white}{\colourone} & $\delta P_{\bar{1}}=\delta P_{\bar{2}}=0$ & $(3,-5)$\tabularnewline
 & $3f$ & $\kappa\cos^{2}\phi$ & \smalldynkin{white}{\colourone}{\colourone} & $\delta P_{\bar{1}}=0$, $\delta P_{\bar{2}}=\delta P_{\bar{3}}$ & $(1,-5)$\tabularnewline
\hline 
Heavy: & $4$ & $\kappa$ & \smalldynkin{\colourone}{\colourtwo}{\colourone} & $\delta P_{\bar{1}}=\tfrac{1}{2}\delta P_{\bar{2}}=\delta P_{\bar{3}}$ & $(1,-1)$\tabularnewline
 & $4f$ & $\kappa$ & \smalldynkin{\colourone}{\colourone}{\colourone} & $\delta P_{\bar{1}}=\delta P_{\bar{2}}=\delta P_{\bar{3}}$ & $(1,-3)$\tabularnewline
\hline 
\end{tabular}\protect\caption[Fake caption without tikz figures.]{List of modes in the $AdS_{3}\times S^{3}\times S^{3}\times S^{1}$
algebraic curve, now including massless fermions. The colouring of
the nodes is $-k_{\ell r}$ with $\smallcirc{\colourone}= 1$ and
$\smallcirc{\colourtwo}=2$. We omit here the right-hand modes $\bar{0}\ldots\bar{4}f$,
for which $k_{\ell\bar{r}}=-k_{\ell r}$. \label{tab:List-of-modes} }
\end{table}

Some further comments:
\begin{itemize}
\item In the literature only the nodes $1$ and $3$ are described momentum-carrying,
and thus it may seem a little puzzling that our $0f$ mode excites
neither of them. However the notion of which nodes carry momentum
depends on the choice of vacuum,%
\footnote{We thank Konstantin Zarembo for explaining this to us. %
} and in general we should define%
\footnote{The sign of $P$ is chosen to match that in \cite{Zarembo:2010yz,Abbott:2012dd};
the factor $1/2\pi\kappa$ is due to the normalisation of $\vec{\kappa}$
in \eqref{eq:delta-vac-p}. %
} 
\begin{align}
P_{\zeta} & =-\frac{1}{2\pi\kappa}\kappa_{\ell}A_{\ell m}P_{m}\label{eq:P-nodes-zeta}\\
 & =-2\sin\phi\sin\zeta\:(P_{1}-P_{\bar{1}})-\left[1-\cos(\phi-\zeta)\right](P_{2}-P_{\bar{2}})-2\cos\phi\cos\zeta\:(P_{3}-P_{\bar{3}}).\nonumber 
\end{align}
At $\zeta=\phi$ this gives the familiar 
\[
P_{\phi}=-2\sin^{2}\phi\:(P_{1}-P_{\bar{1}})-2\cos^{2}\phi\:(P_{3}-P_{\bar{3}}).
\]
At $\zeta=\phi+\pi$, note that it is $4f$ which is does not appear
to be momentum-carrying, while $0f$ carries $2$ units in the expected
reversal of roles. When applied to the perturbations $\delta p$,
this definition gives $\delta P=n/2g\kappa$ for all modes. If we
interpret \eqref{eq:delta-vac-freq-4f} as being a giant magnon dispersion
relation, note that $4h^{2}\sin^{2}(\delta P/2)=n^{2}/\kappa^{2}$
as expected.
\item One may wonder at this point whether there is a special connection
between heavy and massless modes. (The heavy modes are, after all,
composite objects in the same sense as in $AdS_{4}\times CP^{3}$.)
We believe that this is not the case, because there are not just two
supersymmetric pointlike solutions, but four: see figure \ref{fig:modes-of-point-particle-zeta}.
And going for instance from $\zeta=\phi$ to $\zeta=\pi-\phi$ re-organises
$1f\leftrightarrow4f$ and $3f\leftrightarrow0f$ instead of \eqref{eq:reversal-symm-on-modes},
thus mixing light and massless modes instead (and also light and heavy).
In place of \eqref{eq:reversal-symm-q-and-p} we should use 
\[
p''(x)=\begin{pmatrix}p_{1}(x)\\
p_{2}(x)\\
-p_{3}(x)+p_{2}(x)
\end{pmatrix}.
\]
This clearly gives the same re-organisation $1f\leftrightarrow4f$
and $3f\leftrightarrow0f$ for the modes of an arbitrary solution
$p(x)$, not just the point particle. At $\phi=\tfrac{\pi}{4}$ it
reads $q_{1}''=q_{1}$, $q_{3}''=-q_{5}$ and $q_{5}''=-q_{3}$, but
away from this it is more messy in terms of the $q_{i}$ (as we would
expect). 
\item The list of allowed modes in table \ref{tab:List-of-modes} is extra
information not contained in the finite gap equations, usually written
\cite{Babichenko:2009dk,Borsato:2012ss}
\[
2\pi n_{\ell}=-A_{\ell j}\frac{\kappa_{j}x+2\pi m_{j}}{x^{2}-1}+A_{\ell m}\rlap{\;\,--}\int dy\frac{\rho_{m}(y)}{x-y}-A_{\ell k}S_{km}\int\frac{dy}{y^{2}}\frac{\rho_{m}(y)}{x-\tfrac{1}{y}}.
\]
For example, the cut corresponding to the light boson ``$3$'' of
course involves turning on density $\rho_{3}$ alone, with mode number
$n=n_{3}$. But most of the other modes involve turning on several
densities $\rho_{\ell}$ in a correlated way, and further allowing
that only the sum of their $n_{\ell}$ will be an integer: for the
light fermion $3f$ we have $n=n_{2}+n_{3}\in\mathbb{Z}$. 
\item In the worldsheet language we can embed $S_{+}^{3}\in\mathbb{C}^{2}$
by $X=(\sin\theta_{+}e^{i\varphi_{+}},\cos\theta_{+}e^{i\beta_{+}})$,
and similarly $Y=(\sin\theta_{-}e^{i\varphi_{-}},\cos\theta_{-}e^{i\beta_{-}})$.
The effect of \eqref{eq:reversal-symm-q-and-p} is then $X_{i}'=X_{i}^{*}$
and $Y_{i}'=Y_{i}^{*}$, or $\varphi_{\pm}'=-\varphi_{\pm}$, $\beta_{\pm}'=-\beta_{\pm}$,
and this is clearly a symmetry of the action. 
\end{itemize}

\subsection{The Circular String}

It is trivial to integrate the solution \eqref{eq:circular-class-WS}
using \eqref{eq:LS-residue-integrals} and \eqref{eq:LS-residue-integral-u}
to get the following residues: 
\begin{align}
\kLS_{0} & =i2\pi\kappa & \kLS_{1} & =-2\pi\omega_{+}/\cos\phi & \kLS_{2} & =-2\pi\omega_{-}/\sin\phi & \kLS_{3} & =-2\pi\omega_{u}\label{eq:circular-residues}\\
\mLS_{0} & =im_{0}=0 & \mLS_{1} & =-m_{+}/\cos\phi & \mLS_{2} & =-m_{-}/\sin\phi & \mLS_{3} & =-m_{u}.\nonumber 
\end{align}
The traditional Virasoro condition \eqref{eq:AC-vir-traditional}
gives (setting $D=3$ there to include $\pLS_{3}$) 
\[
-(\kappa\pm m_{0})^{2}+\frac{(\omega_{+}\pm m_{+})^{2}}{\cos^{2}\phi}+\frac{(\omega_{-}\pm m_{-})^{2}}{\sin^{2}\phi}+(\omega_{u}\pm m_{u})^{2}=0
\]
agreeing with the worldsheet one \eqref{eq:circular-WS-vir}. Again
we write $m_{0}$ here just to allow for $\delta m_{0}$ in \eqref{eq:circle-linearised-virasoro}
below. 

Mapping this into $AdS_{3}\times S^{3}\times S^{3}$ using \eqref{eq:my-map-p-from-pLS},
we ignore $\pLS_{3}$ (describing the $S^{1}$ factor) for now. The
momentum carried is $P_{\bar{1}}=-m_{-}\pi/\sin^{2}\phi$ and $P_{\bar{3}}=-m_{+}\pi/\cos^{2}\phi$,
which can be combined using \eqref{eq:P-nodes-zeta} to give total
momentum 
\[
P_{\zeta}=2\pi\left(m_{+}\frac{\cos\zeta}{\cos\phi}+m_{-}\frac{\sin\zeta}{\sin\phi}\right).
\]
If we use the momentum assignments from the $\zeta=\phi$ vacuum (or
any supersymmetric vacuum) then $P_{\zeta}$ is a multiple of $2\pi$
(as expected for a closed string) whenever $m_{\pm}\in\mathbb{Z}$. 

To construct modes for this solution, we need two extensions to what
we have done above. First, we should allow the residues of the poles
at $x=\pm1$ to vary independently, but still subject to the Virasoro
constraint. This means that we allow the windings $m_{\ell}$ to vary.
Second, we find it necessary to demand that the resulting $\delta P_{\ell}$
belonging to nodes which are excited are all equal, and the others
zero (and the same for the $\delta P_{\bar{\ell}}$ of the inverted
sheets).

To be explicit, the perturbation of the quasimomenta is 
\begin{equation}
\delta p_{\ell}(x)=k_{\ell}\left[\frac{\alpha(y)}{x-y}+\frac{1}{2}\frac{\alpha(y)}{y}\right]+S_{\ell m}k_{m}\left[\frac{\alpha(y)}{\frac{1}{x}-y}+\frac{1}{2}\frac{\alpha(y)}{y}\right]+\sum_{v}\frac{\partial p_{\ell}(x)}{\partial v}\delta v\label{eq:circle-delta-p_ell}
\end{equation}
where $v\in\{\kappa,\omega_{\pm},m_{0},m_{\pm}\}$,%
\footnote{Note that we do not vary $\omega_{u}$ and $m_{u}$, because we do
not have a condition at infinity on $\pLS_{3}$. While we write the
variables shown in \eqref{eq:circular-residues}, we could clearly
use $v\in\{\kappa_{\ell},m_{\ell};\ell=1,2,3\}$ instead. %
} subject to the following linearised Virasoro constraints: 
\begin{equation}
\sum_{v}\frac{\partial V^{\text{diag}}}{\partial v}\delta v=0,\qquad\sum_{v}\frac{\partial V^{\text{off}}}{\partial v}\delta v=0.\label{eq:circle-linearised-virasoro}
\end{equation}
We impose the following condition at infinity 
\begin{equation}
\delta p_{\ell}(x)\to k_{\ell}\:\delta P_{\text{left}}-k_{\bar{\ell}}\:\delta P_{\text{right}}+\frac{1}{2gx}\left(k_{\ell}+D_{\ell}\delta\Delta\right)+\bigodiv{x^{2}}\label{eq:circle-cond-at-infty}
\end{equation}
and solve for $\delta v,\delta P_{\text{left}},\delta P_{\text{right}}$
and $\delta\Delta=\Omega_{r}(y)$. The on-shell frequency is found
by solving 
\[
2\pi n=-k_{\ell}A_{\ell m}p_{m}(y_{n})
\]
for $y_{n}$ and evaluating $\Omega(y_{n})$. 

We can write all the results in a compact form, in terms of two numbers
$S_{r}$ $M_{r}$:
\begin{equation}
\Omega_{r}(y)=\frac{1}{y^{2}-1}\frac{S_{r}^{2}-M_{r}^{2}}{\kappa S_{r}},\qquad2\pi n_{r}=2\pi\frac{S_{r}y+M_{r}}{y^{2}-1}\label{eq:circle-S-M-omega-and-n}
\end{equation}
for left-hand modes, and $2\pi n_{r}=2\pi\frac{S_{r}y+M_{r}y^{2}}{y^{2}-1}$
for right-hand (barred) modes. The on-shell frequency is then
\begin{equation}
\negthickspace\negthickspace\Omega_{r}(y_{n})=-\frac{S_{r}^{2}-M_{r}^{2}}{2\kappa S_{r}}-\left(n\pm\frac{M_{r}}{2}\right)\frac{M_{r}}{\kappa S_{r}}+\frac{1}{\kappa}\sqrt{\left(n\pm\frac{M_{r}}{2}\right)^{2}+\frac{S_{r}^{2}-M_{r}^{2}}{4}},\qquad\begin{small}r\in\negthickspace\begin{array}{l}
\text{left}\\
\text{right.}
\end{array}\end{small}\label{eq:circular-on-shell-S-M}
\end{equation}
The coefficients here are the same left and right ($S_{r}=S_{\bar{r}}$
and $M_{r}=M_{\bar{r}}$) and are given by 
\begin{equation}
\begin{aligned}S_{0} & =2s_{0} & M_{0} & =0 & S_{0f} & =\kappa-\omega_{+}-\omega_{-} & M_{0f} & =-m_{+}-m_{-}\\
S_{1} & =2\omega_{-} & M_{1} & =2m_{-} & S_{1f} & =\kappa-\omega_{+}+\omega_{-} & M_{1f} & =-m_{+}+m_{-}\\
S_{3} & =2\omega_{+}\quad & M_{3} & =2m_{+}\qquad & S_{3f} & =\kappa+\omega_{+}-\omega_{-}\quad & M_{3f} & =m_{+}-m_{-}\\
S_{4} & =2\kappa & M_{4} & =0 & S_{4f} & =\kappa+\omega_{+}+\omega_{-} & M_{4f} & =m_{+}+m_{-}.
\end{aligned}
\label{eq:circular-bigS-bigR-table}
\end{equation}
The masses $s_{r}^{2}=(S_{r}^{2}-M_{r}^{2})/4$ are identical to those
from the worldsheet calculation, \eqref{eq:WS-masses-bosons-circular}
and \eqref{eq:WS-masses-fermions-circular}. Some comments follow:
\begin{itemize}
\item Compared to the worldsheet results, the frequency \eqref{eq:circular-on-shell-S-M}
displays a shift in $n$ and a shift in energy. In the simplest case
\eqref{eq:circular-simple-case} $M_{r}/2\in\mathbb{Z}$ and thus
the shift in $n$ will not matter, but in general it may be a half-integer,
in which case according to \cite{Mikhaylov:2010ib} we should trust
this not the worldsheet one. The term linear in $n$ vanishes in the
sum. For the shifts in energy, note that we are still missing the
massless bosons, and see discussion in \cite{Gromov:2007aq,Abbott:2010yb}
.
\item One surprising feature is that the ``efficient'' method \cite{Gromov:2008ec}
of constructing heavy modes off-shell by addition given by \cite{Abbott:2012dd}
fails here. The rules were 
\begin{equation}
1f+3f=4,\qquad1+3f=1f+3=4f\label{eq:efficient-rules-2012}
\end{equation}
and these still hold at the level of $k_{\ell}$ of course (as is
easily seen from table \ref{tab:List-of-modes}), but not at the level
of $\Omega_{r}(y)$.%
\footnote{These rules do hold for the point particle case above, which has no
winding and thus has $\delta m=0$. We imposed this in \eqref{eq:delta-vac-delta-kappa}
but it is still true using the more liberal \eqref{eq:circle-linearised-virasoro}
for this solution.%
}  We can observe however that \eqref{eq:efficient-rules-2012} are
not invariant under \eqref{eq:reversal-symm-on-modes}: the second
equation becomes 
\[
-1+1f=3f-3=0f.
\]

\item We discuss massless bosons in section \ref{sub:Massless-Bosons?}
below. 
\end{itemize}

\subsection{The Folded String}

Because the solution \eqref{eq:folded-class-WS} also lives in $\mathbb{R}\times(S^{1})^{3}$
it is also described by only poles at $x=\pm1$. We can use \eqref{eq:LS-residue-integrals}
and \eqref{eq:LS-residue-integral-u} to integrate and find their
residues; notice that $A$ does not appear: 
\[
\kLS_{0}=i2\pi\kappa,\qquad\kLS_{1}=-2\pi\omega\,\cos\phi,\qquad\kLS_{2}=-2\pi\omega\,\sin\phi,\qquad\kLS_{3}=0.
\]
Since \eqref{eq:circular-residues} is the most general set of residues,
we can describe this as a special case 
\begin{equation}
\omega_{+}=\omega\:\cos^{2}\phi,\qquad\omega_{-}=\omega\:\sin^{2}\phi,\qquad m_{\pm}=0,\qquad\omega_{u}=m_{u}=0.\label{eq:folded-AC-special-case}
\end{equation}
An important difference from the circular string is that the traditional
Virasoro constraint \eqref{eq:AC-vir-traditional} is not obeyed.
It gives $\kappa^{2}=\omega^{2}$ here, contradicting the worldsheet
one which gives $\kappa^{2}=A^{2}\nu^{2}+\omega^{2}$, the physical
condition that the cusps are lightlike. Thanks to \cite{Lloyd:2013wza}
we understand that this is not a problem: \eqref{eq:AC-vir-traditional}
is too strict, and their GRC \eqref{eq:AC-vir-new-GRC} allows for
the residues seen here. 

When we calculate modes of this solution, however, we demand that
$\delta p(x)$ still obeys the linearised condition \eqref{eq:circle-linearised-virasoro}
derived from \eqref{eq:circular-WS-vir}. Note that this was the only
point at which the Virasoro constraints entered the mode computation
for the circular string: at no point did we use the fact that the
classical $p(x)$ obeys \eqref{eq:circular-WS-vir}. And thus nothing
changes in our calculation of the mode frequencies. Substituting \eqref{eq:folded-AC-special-case}
into the circular string's mode masses, we obtain 
\begin{align}
 &  & s_{0f} & =\tfrac{1}{2}(\kappa-\omega)\nonumber \\
s_{1} & =\omega\:\cos^{2}\phi & s_{1f} & =\tfrac{1}{2}(\kappa+\omega\,\cos2\phi)\displaybreak[0]\label{eq:folded-AC-masses}\\
s_{3} & =\omega\:\sin^{2}\phi & s_{3f} & =\tfrac{1}{2}(\kappa-\omega\,\cos2\phi)\nonumber \\
s_{4} & =\kappa & s_{4f} & =\tfrac{1}{2}(\kappa+\omega).\nonumber 
\end{align}
These match what we got in the worldsheet theory, sections \ref{sub:Bosonic-Modes-of-Folded-String}
and \ref{sub:Fermionic-Modes-of-Folded-String}. 

Using the linearised Virasoro condition \eqref{eq:circle-linearised-virasoro}
for these modes is justified by our worldsheet calculation. There
we showed that, for both this solution and the circular string, only
the massless bosons produce any change to any $\delta V_{l}$, the
different factors' contributions to the Virasoro constraint. Thus
for all the modes studied here, the change to the integrated form
\eqref{eq:AC-vir-traditional} will be zero. (We believe this will
be true for any classical solution.)

The classical solution \eqref{eq:class-LS} can be treated as a special
case of the circular string in exactly the same way. Its residues
$\kLS_{l}$, $\mLS_{l}$ were written down in (4.49) of \cite{Lloyd:2013wza},
and can be plugged into the masses $s_{r}^{2}=(S_{r}^{2}-M_{r}^{2})/4$
after using \eqref{eq:circular-residues}. Let us write just the special
case $a=\tilde{a}$, $\nu=\tilde{\nu}$ which has zero winding, $\mLS_{l}=0$,
and residues 
\begin{equation}
\kLS_{1}=-2\pi\omega\,\cos\phi,\qquad\kLS_{2}=-2\pi\omega\,\sin\phi,\qquad\omega=\frac{2\kappa}{\pi}\, E\Big(\frac{\pi}{2}\Big\vert\frac{4a^{2}\nu^{2}}{\kappa^{2}}\Big).\label{eq:LS-special-case}
\end{equation}
Then its mode masses are given by \eqref{eq:folded-AC-masses} with
this $\omega$. Notice in particular that none of the fermionic modes
of this are truly massless. In the limit $a\to0$, in which this should
approach the BMN solution, we have $E(\frac{\pi}{2}\vert m)=\frac{\pi}{2}-\frac{\pi}{8}m+\bigo{m^{2}}$
and thus 
\begin{equation}
\omega=\kappa-\frac{a^{2}\nu^{2}}{\kappa}+\bigo{a^{4}}\qquad\Rightarrow\qquad\begin{aligned}s_{0f} & =0+\frac{a^{2}\nu^{2}}{2\kappa}+\bigo{a^{4}}\\
s_{1f} & =\kappa\:\cos^{2}\phi-\frac{a^{2}\nu^{2}}{2\kappa}\cos2\phi+\bigo{a^{4}},\;\mbox{ etc.}
\end{aligned}
\label{eq:LS-mass-in-limit}
\end{equation}
While finding the modes of the exact macroscopic solution \eqref{eq:class-LS}
in the worldsheet language seems hard, perhaps it would be possible
to calculate these corrections to the BMN masses. This would be an
interesting check of our methods.

\subsection{Massless Bosons?\label{sub:Massless-Bosons?}}

We showed above how to incorporate the massless fermions into the
algebraic curve structure. Naturally we wonder if something similar
can be done for the massless bosons, to capture all $8+8$ modes. 

In the worldsheet language we know these modes exactly --- they are
fluctuations in directions in target space for which $\mathcal{L}_{2}$
is exactly the free Lagrangian $\partial_{\mu}\tilde{X}\,\partial^{\mu}\tilde{X}$
with no mass term, and the solutions are plane waves. These are precisely
the directions within $\mathbb{R}\times(S^{1})^{3}$, for which we
can use \eqref{eq:LS-residue-integrals} to work out the algebraic
curve. Since these equations are linear, we can also work out the
change to the algebraic curve, regardless of the classical solution
being perturbed. And the answer is simply zero. The pessimistic answer
is thus that they are invisible to this formalism. 

In table \eqref{eq:circular-bigS-bigR-table} we were a little more
optimistic, and included $S_{0}=2s_{0}$ without having solved for
it. We do this on the grounds that we believe \eqref{eq:circle-S-M-omega-and-n}
and \eqref{eq:circular-on-shell-S-M} should, in the limit $s_{0}\to0$,
correctly describe this mode. To understand the limit we look at the
$4f$ mode near to $s_{4f}=0$ (i.e. near to $\cos(\phi-\zeta)=-1$).
The energy \eqref{eq:delta-vac-freq-4f} has the expected finite limit
$\omega_{n}=\left|n\right|/\kappa$, but this arises from dividing
zero by zero: the off-shell frequency $\Omega(y)$ \eqref{eq:delta-vac-Omega-4f}
appears to go to zero, but (holding $n$ fixed) the position of the
pole $y_{n}$ from \eqref{eq:delta-vac-2pin-4f} also approaches $1$.
The bosonic modes behave the same way, for instance the $3$ mode
near to $\zeta=\pi/2$ (i.e. near to $J_{+}=0$). 

Perhaps the same idea of everything converging on $x=1$ applies also
to macroscopic classical solutions: moving in ``massive'' directions
these would be described by one- or two-cut resolvents, but the ``massless''
versions studied here are described by just the poles. 

Finally note that we have also largely ignored the $S^{1}$ direction
$u$, which gives one of the massless bosons in the worldsheet picture.
This appears to be correct in the sense that none of the worldsheet
frequencies depend on the classical solution's $\omega_{u}$ and $m_{u}$.
We constructed $\pLS_{3}$ in \eqref{eq:LS-residue-integral-u}, and
could write $p_{4}=-\pLS_{3}$ and $A_{\text{all}}=(A_{\text{left}}\oplus1)\otimes1_{2\times2}$.
Then provided no modes turn on this node (i.e. $k_{4}=0$ always)
nothing will change in our calculations.%
\footnote{We can then allow $v$ in \eqref{eq:circle-linearised-virasoro} to
include $\omega_{u},m_{u}$. %
}

\section{Energy Corrections\label{sec:Energy-Corrections}}

The last two sections developed two ways to calculate mode frequencies
of the ``macroscopic massless'' spinning strings that we are studying.
The main reason to do so at all is to work out the one-loop correction
to the energy, by adding these frequencies up. 

For frequencies of the form $\omega_{n}=\tfrac{1}{\kappa}\sqrt{(n+m)^{2}+s^{2}}$
(in which we allow some shift of the mode number by $m=\pm M_{r}/2$)
the one-loop energy correction is 
\begin{equation}
\delta E=\sum_{n=-N}^{N}e(n)=\sum_{r}(-1)^{F}\sum_{n=-N}^{N}\frac{1}{2\kappa}\sqrt{(n+m_{r})^{2}+s_{r}^{2}}\label{eq:defn-e(n)}
\end{equation}
(defining $e(n)$ for use below). The simplest way to evaluate this
sum is to approximate it with an integral: provided that $\sum_{r}(-1)^{F}=0$
and $\sum_{r}(-1)^{F}s_{r}^{2}=0$, to ensure (respectively) that
the quadratic and logarithmic divergences cancel, we have
\begin{align}
\delta E & \approx\frac{\kappa}{2}\sum_{r}(-1)^{F}\int_{-N/\kappa}^{N/\kappa}dz\:\sqrt{\Big(z+\frac{m_{r}}{\kappa}\Big)^{2}+\Big(\frac{s_{r}}{\kappa}\Big)^{2}},\qquad z=n/\kappa\nonumber \\
 & =\frac{1}{2\kappa}\sum_{r}(-1)^{F}\left[-s_{r}^{2}\log s_{r}+m_{r}^{2}\right].\label{eq:simple-dE-s-log-s}
\end{align}

Applying this method to the folded string, using the masses \eqref{eq:folded-AC-masses},
it simplifies when we take the string to be short, $A^{2}\ll\kappa^{2}$.
This is equivalent to taking $\omega=\cJ$ large (compared to $\cS$),
and for the sake of familiarity we write all energy corrections in
terms of these angular momenta. At $\phi=\tfrac{\pi}{4}$ we get 
\begin{align*}
\delta E & =-\frac{\nu\cS}{\mbox{\ensuremath{\cJ}}}\log2+\frac{\nu^{2}\cS^{2}}{8\cJ^{3}}\left[-3+6\log2-4\log\cJ+\log\nu(\cS)\right]\\
 & \qquad+\frac{\nu^{3}\cS^{3}}{8\cJ^{5}}\left[1-2\log2+2\log\cJ-\log\nu(\cS)\right]+\bigodiv{\cJ^{5}}.
\end{align*}
However this is almost certainly not what we want to do. For the case
of $sl(2)$ circular strings in $AdS_{5}\times S^{5}$, agreement
with the Bethe equations was seen by expanding $e(n)$ in large $\cJ$
first. Naively this leads to a divergent result, but Beisert and Tseytlin
\cite{Beisert:2005cw} showed how to re-sum the divergent terms. We
now adapt what they did to the frequencies seen here, still focusing
on the folded string.

\subsection{Adapting the Beisert--Tseytlin procedure}

\newcommand{\esum}{\smash{e^\text{sum}}}
\newcommand{\esumsing}{\smash{e^\text{sum}_\text{sing}}}
\newcommand{\esumreg}{\smash{e^\text{sum}_\text{reg}}}
\newcommand{\eint}{\smash{e^\text{int}}}
\newcommand{\eintsing}{\smash{e^\text{int}_\text{sing}}}
\newcommand{\eintreg}{\smash{e^\text{int}_\text{reg}}}
\newcommand{\muh}{\hat{\mu}}The procedure given by \cite{Beisert:2005cw} took $e(n)=\sum_{r}(-1)^{F_{r}}\omega_{n}^{r}/2$
and divided it up as follows: After expanding in $\mathcal{J}\gg1$
at fixed $n$ to get $\esum(n)$, they defined 
\begin{align*}
\esumsing(n) & =\mbox{terms in }\esum(n)\mbox{ which give a divergence in the sum to }n=\infty,\\
\esumreg(n) & =\esum(n)-\esumsing(n).
\end{align*}
Then they wrote $n=\mathcal{J}x$ and expanded $e(\mathcal{J}x)$
in $\mathcal{J}\gg1$ at fixed $x$ to get $\eint(x)$, and similarly
defined 
\begin{align*}
\eintsing(x) & =\mbox{terms in }\eint(x)\mbox{ which give a divergence in the integral to }x=0,\\
\eintreg(x) & =\eint(x)-\eintsing(x).
\end{align*}
It was observed that $\esumreg(n)=\eintsing(n/\mathcal{J})$ and similarly
$\esumsing(n)=\eintreg(n/\cJ$), allowing them to exchange the singular
part of a sum on $n$ for the regular part of an integral on $x$.
The total energy correction was then 
\begin{equation}
\delta E=\delta E^{\text{sum}}+\delta E^{\text{int}}=\sum_{n=-\infty}^{\infty}\esumreg(n)+\int_{-\infty}^{\infty}\mathcal{J}dx\:\eintreg(x).\label{eq:dE-sumreg+intreg}
\end{equation}
Here $\delta E^{\text{sum}}$ was the analytic part, containing only
even powers of $\mathcal{J}=J/\sqrt{\lambda}$, and $\delta E^{\text{int}}$
was the non-analytic part, containing odd powers of $\mathcal{J}$.
The necessity of reproducing such non-analytic terms is what led \cite{Beisert:2005cw}
to introduce the one-loop dressing phase in the Bethe equations. 

Applying this procedure here, we find that $\eintsing(x)$ has odd
powers including $1/x$ at $x=0$, which lead to terms in $\eintreg(x)$
which behave like $1/x$ at $x=\infty$, giving a logarithmically
divergent answer. This problem does not arise in \cite{Beisert:2005cw},
nor in \cite{Beccaria:2012kb}, where the divergent terms start only
at order $1/x^{2}$ and at order $n^{2}$. 

Let us consider the following modification of $e_{\text{sing}}$:
\begin{align}
\esumsing(n) & =\left[\mbox{terms in }\esum(n)\mbox{ which go like }n^{0}\mbox{ or faster as }n\to\infty\right]\nonumber \\
 & \qquad\qquad+\mu\left[\mbox{terms in }\esum(n)\mbox{ which go like }n^{-1}\mbox{ as }n\to\infty\right],\label{eq:dE-sing-mu-nu}\\
\eintsing(x) & =\left[\mbox{terms in }\eint(x)\mbox{ which go like }x^{-2}\mbox{ or faster as }x\to0\right]\nonumber \\
 & \qquad\qquad+\muh\left[\mbox{terms in }\eint(x)\mbox{ which go like }x^{-1}\mbox{ as }x\to0\right].\nonumber 
\end{align}
Clearly $\mu=\muh=1$ is the original procedure as described above,
but the parameters $\mu,\muh$ will let us control the new logarithmic
divergences, without altering any results of \cite{Beisert:2005cw,Beccaria:2012kb}.
Introduce three explicit cutoffs as follows: 
\begin{equation}
\delta E^{\text{sum}}=2\sum_{n=1}^{N}\esumreg(n),\qquad\delta E^{\text{int}}=2\int_{\epsilon}^{\Lambda}\mathcal{J}dx\:\eintreg(x).\label{eq:dE-int-sum-with-cutoffs}
\end{equation}
Treating the folded spinning string (using the masses \eqref{eq:folded-AC-masses},
and writing $\cSn=\nu\,\cS$) we obtain 
\begin{align}
\delta E^{\text{sum}} & =\mbox{finite}+(\mu-1)\left(\log N+\gamma_{E}\vphantom{1^{1^{1}}}\right)\left[\frac{\cSn^{2}}{4\cJ^{3}}-\frac{\cSn^{3}}{2\cJ^{5}}+\frac{15\cSn^{4}}{16\cJ^{7}}+\bigodiv{\cJ^{9}}\right]\nonumber \\
\delta E^{\text{int}} & =\mbox{finite}+\left[\muh\log\Lambda+(1-\muh)\log\epsilon\vphantom{1^{1^{1}}}\right]\left[\frac{\cSn^{2}}{4\cJ^{3}}-\frac{\cSn^{3}}{2\cJ^{5}}+\frac{15\cSn^{4}}{16\cJ^{7}}+\ldots\right].\label{eq:dE-divergent-terms}
\end{align}
To match the upper cutoff of the sum with the lower cutoff of the
integral, we want $N=\mathcal{J}\epsilon$. To cancel all three of
the divergences,%
\footnote{It seems convenient to absorb the terms in $\gamma_{E}=0.577\ldots$
here, since they are of the same form.%
} we need
\[
N=e^{-\gamma_{E}}\:\Lambda^{\frac{\muh}{1-\mu}}\:\epsilon^{\frac{1-\muh}{1-\mu}}.
\]
At $\mu=\muh=\tfrac{1}{2}$, both of these conditions are solved by
$N=\sqrt{\cJ}$, $\epsilon=1/\sqrt{\cJ}$, $\Lambda=e^{\gamma_{E}}\cJ$. 

The other constraint on $\mu,\muh$ is that we must again have $\esumreg(n)=\eintsing(n/\mathcal{J})$,
in order to omit $\eintsing$ and $\esumsing$ from \eqref{eq:dE-sumreg+intreg}.
The first few terms of these are%
\footnote{The terms shown come from expanding $\esum$ and $\eint$ up to $1/\cJ^{10}$.
The $1/n^{5}$ term missing from $\eintsing(n/\cJ)$ is order $1/\cJ^{12}$
in $\eint(x)$ and thus not yet visible.%
} 
\begin{align}
\esumreg(n) & =\frac{-\cSn^{2}\frac{(1-\mu)}{8n}}{\cJ^{3}}+\frac{\cSn^{3}\frac{(1-\mu)}{4n}+\cSn^{4}\frac{1}{128n^{3}}}{\cJ^{5}}-\frac{\cSn^{4}\frac{15(1-\mu)}{32n}+\cSn^{5}\frac{3}{128n^{3}}+\cSn^{6}\frac{1}{1024n^{5}}}{\cJ^{7}}+\bigodiv{\cJ^{7}}\nonumber \\
\eintsing(n/\mathcal{J}) & =\frac{-\cSn^{2}\frac{\muh}{8n}}{\cJ^{3}}+\frac{\cSn^{3}\frac{\muh}{4n}+\cSn^{4}\frac{1}{128n^{3}}}{\cJ^{5}}-\frac{\cSn^{4}\frac{15\muh}{32n}+\cSn^{5}\frac{3}{128n^{3}}+\ldots}{\cJ^{7}}+\ldots\label{eq:dE-magic-cancellation}
\end{align}
and clearly $\mu=\muh=\tfrac{1}{2}$ gives agreement. 

After cancelling these divergences, the finite parts are 
\begin{align}
\delta E^{\text{sum}} & =\frac{\cSn^{4}\:\zeta(3)}{64\cJ^{5}}-\frac{\cSn^{5}\:24\zeta(3)+\cSn^{6}\:\zeta(5)}{512\cJ^{7}}\nonumber \\
 & \qquad+\frac{\cSn^{6}\,\zeta(3)\,7\cdot2^{-6}+\cSn^{7}\,\zeta(5)\,2^{-7}+\cSn^{8}\,\zeta(7)\,5\cdot2^{-14}}{\cJ^{9}}+\bigodiv{\cJ^{11}}\nonumber \\
\shortintertext{and}\delta E^{\text{int}} & =\frac{\mbox{\ensuremath{\cS}}}{\mbox{\ensuremath{\cJ}}}\left[\alpha\log\alpha+(1-\alpha)\log(1-\alpha)\vphantom{1^{1^{1}}}\right]\nonumber \\
 & \qquad+\frac{\cS^{2}}{4\cJ^{3}}\left[(1-6\alpha)\log\alpha+(6\alpha-5)\log(1-\alpha)+\log2-1\vphantom{1^{1^{1}}}\right]+\bigodiv{\cJ^{5}}.\label{eq:dE-int-finite}
\end{align}
Note that there appears to be no consistent pattern of even and odd
powers, like that for the analytic / non-analytic distinction. Note
also that only $\delta E^{\text{int}}$ depends on $\phi$, which
we write here as $\alpha=\cos^{2}\phi$. At $\phi=\pi/4$ our result
is simpler,
\[
\delta E^{\text{int}}=-\frac{\cSn\log2}{\mbox{\ensuremath{\cJ}}}+\frac{\cSn^{2}(5\log2-1)}{4\cJ^{3}}-\frac{\cSn^{3}(8\log2-1)}{4\cJ^{5}}+\frac{\cSn^{4}(660\log2-113)}{192\cJ^{7}}+\bigodiv{\cJ^{9}}.
\]

Integrability normally gives an expansion in the Bethe coupling $h$,
rather than $\sqrt{\lambda}$. These are related here by \cite{Sundin:2012gc,Abbott:2012dd,Beccaria:2012kb}%
\footnote{This assumes we are using a cutoff on energy or mode number, rather
than in the spectral plane. In the similar relation for $AdS_{4}\times CP^{3}$
\cite{Gromov:2008fy,McLoughlin:2008he,Abbott:2010yb,Astolfi:2011ju,Abbott:2011xp},
this was ultimately understood to be the correct choice \cite{Gromov:2014eha,Bianchi:2014ada}.%
} 
\begin{equation}
h(\lambda)=\frac{\sqrt{\lambda}}{2\pi}+c+\ldots=\frac{\sqrt{\lambda}}{2\pi}+\frac{\alpha\log\alpha+(1-\alpha)\log(1-\alpha)}{2\pi}+\mathcal{O}\Big(\frac{1}{\sqrt{\lambda}}\Big).\label{eq:h-and-c}
\end{equation}
If we regard the classical energy \eqref{eq:folded-WS-charge-rel}
as the zeroth term in an expansion in $h$, writing $\cJh=J/2\pi h$
(so that $\cJh=\cJ$ classically), then expanding in $\sqrt{\lambda}$
gives the following terms: 
\begin{align*}
\Delta_{h=\infty} & =2\pi h\sqrt{\cJh^{2}+2\nu\cS_{h}}\\
 & =\sqrt{\lambda}\sqrt{\cJ^{2}+2\cSn}+\frac{2\pi c\,\cSn}{2\cJ\sqrt{\cJ^{2}+2\cSn}}+\bigodiv{\sqrt{\lambda}}\\
 & =\sqrt{\lambda}\sqrt{\cJ^{2}+2\cSn}+2\pi c\left[\frac{\cSn}{\mbox{\ensuremath{\cJ}}}-\frac{\cSn^{2}}{\cJ^{3}}+\frac{3\cSn^{3}}{2\cJ^{5}}+\bigodiv{\cJ^{7}}\right]+\bigodiv{\sqrt{\lambda}}.
\end{align*}
The $\cS/\cJ$ term here is exactly the first term seen in \eqref{eq:dE-int-finite}.
But starting with the $1/\cJ^{3}$ term there are genuine $1/h$ corrections,
i.e. $\bigo{h^{0}}$ terms. The first few terms are 
\[
\delta\Delta_{h}=\frac{\cS^{2}}{4\cJ^{3}}\left(L-1\right)-\frac{\cS^{3}}{2\cJ^{5}}\left(L-\frac{5}{12}-\frac{1}{3\sin2\phi}\right)+\frac{15\cS^{4}}{16\cJ^{7}}\left(L-\frac{5}{36}-\frac{2}{5\sin2\phi}-\frac{4}{45\sin^{2}2\phi}\right)+\ldots
\]
where $L=\log2+(1-2\alpha)\log\alpha-(1-2\alpha)\log(1-\alpha)=\log2+\cos2\phi\log(\tan^{2}\phi$),
and obviously $\sin2\phi=4\alpha(1-\alpha)$.

\section{Conclusions\label{sec:Conclusions}}

The results of this paper are as follows:
\begin{itemize}
\item We have introduced some classical string solutions which we think
can be interpreted as macroscopic excitations of the massless modes
of the BMN string. To describe one of these, the folded string, in
the algebraic curve we must use \cite{Lloyd:2013wza}'s general residue
condition, but the solution here presented is itself much simpler
than their example requiring this. For this property it is necessary
that contributions to Virasoro from different factors of the target
space are not all constant. 
\item We have shown how to calculate the fermionic ``$0f$'' mode frequencies
using the algebraic curve for the first time. This calculation uses
only (a linearised form of) the traditional Virasoro constraint, and
we discussed why this is so. For all the classical solutions considered
here (except BMN) these modes are in fact no longer massless, and
their masses agree with those from worldsheet computations. (By contrast
the bosonic massless modes are always trivial.) 
\item We learned how to do this by studying the reversal symmetry $\omega\to-\omega$
which acts on fermion masses $s\to1-s$. This has not been discussed
in the literature, but it uses the vacuum solutions $\vec{\kappa}$
previously discarded as spurious, and here interpreted as non-BMN
pointlike strings. We can see the same symmetry in $CP^{3}$ and $S^{5}$
cases, but it is not interesting in the absence of massless modes.
Here it is part of a continuous transformation controlled by $\zeta$,
which while not a symmetry, teaches us about the massless limit. 
\item We discovered that the folded string is a special case of circular
string, as far as modes are concerned. We view this as the distinction
between macroscopic one- and two-cut solutions disappearing as these
cuts converge onto the poles at $x=\pm1$, just as the microscopic
cuts which describe the modes converge onto the poles as the mode
becomes massless. The same agreement of frequencies is seen in the
worldsheet language, up to the fact that the modes appear in a slightly
strange gauge. (Perhaps this is a larger example of the frequency
shifts seen for instance in \cite{Gromov:2007aq,Mikhaylov:2010ib}.)
\item We calculated energy corrections $\delta E^{\text{sum}}$ and $\delta E^{\text{int}}$,
which for massive spinning strings would be the analytic and non-analytic
terms. The division between these two types of terms comes from a
version of the Beisert--Tseytlin re-summing procedure, modified here
to deal with logarithmic divergences. We see some $\cos^{2}\phi$
dependence in $\delta E^{\text{int}}$ even when expanding in $h(\lambda)$. 
\end{itemize}
There are many interesting open directions from here:
\begin{itemize}
\item Most obviously, the same reversal idea will allow us to describe in
the algebraic curve the four fermions in $AdS_{3}\times S^{3}\times T^{4}$
which are massless for BMN (sometimes called the non-coset fermions).
This is the topic of a forthcoming paper. Similar things can no doubt
be done for $AdS_{2}\times S^{2}\times T^{6}$ \cite{Sorokin:2011rr,Abbott:2013kka,Hoare:2014kma},
and for backgrounds with mixed RR and NS-NS flux, which are the topic
of much recent interest \cite{Hoare:2013pma,Hoare:2013lja,Wulff:2014kja,David:2014qta,Babichenko:2014yaa,Hernandez:2014eta,Lloyd:2014bsa,Sundin:2014ema,Stepanchuk:2014kza}. 
\item The list of modes in table \ref{tab:List-of-modes} is slightly ad-hoc,
constructed along the lines of earlier examples (such as table \ref{tab:List-of-modes-CP3})
so as to match the BMN spectrum \cite{Abbott:2012dd}, and now the
$\zeta$-vacuum spectrum. Nevertheless it contains important extra
information not present in the finite-gap equations, which follow
directly from $A,S,\vec{\kappa}$. It should be possible to understand
this more rigorously from representation theory.\medskip  \\
Such an understanding might also point out exactly how to deal with
the massless bosons. Our approach here is to view the algebraic curve
as just a tool for calculating mode frequencies, and in this view
there is no need to think about them, since they always have $\omega_{n}=\left|n\right|/\kappa$. 
\item The ultimate goal of studying these spinning strings is to make contact
with the quantum integrable spin-chain picture. In the $T^{4}$ case
this means the S-matrix of \cite{Borsato:2014exa,Borsato:2014hja}
or rather the associated Bethe equations. This S-matrix contains dressing
phases for the massless sector which are at present unknown. \medskip 
\\
However the comparison can't be exactly along the lines of what was
done for $su(2)$ and $sl(2)$ spinning strings \cite{Beisert:2005cw,Hernandez:2006tk},
as there is no meaningful resolvent here. Our discussion of the massless
limit of the modes indicates why: the cuts have collapsed into the
poles at $x=\pm1$. 
\item We observe that several quite different strings can have the same
algebraic curve description: for example \eqref{eq:folded-AC-special-case}
and \eqref{eq:LS-special-case} can easily be made to co-incide exactly.
Some related ideas were explored by \cite{Janik:2012ws}, see also
\cite{Ryang:2012uf,Dekel:2013dy}. 
\item Finally we note that some issues to do with calculating the modes
of folded strings (and other non-homogeneous classical solutions)
were recently encountered by \cite{Forini:2014kza}. Our folded string
is a simpler case, but their techniques may be necessary for more
general solutions. 
\end{itemize}

\subsection*{Acknowledgements}

Chrysostomos Kalousios collaborated on related earlier work. We thank
Romuald Janik, Antal Jevicki, Robert de Mello Koch, Olof Ohlsson Sax,
Alessandro Sfondrini, Bogdan Stefa\'{n}ski, Alessandro Torrielli
and Konstantin Zarembo for helpful discussions. 

Michael was supported by the UCT University Research Council, the
Claude Leon Foundation, and an NRF Innovation Fellowship. In\^{e}s
was partially supported by the FCT--Portugal fellowship SFRH/BPD/69696/2010
and by the NCN grant 2012/06/A/ST2/00396. We thank CERN for hospitality.

\appendix

\section{Reversal Symmetry for $AdS_{4}\times CP^{3}$\label{sec:Reversal-Symmetry-and-AdS4xCP3}}

Above we note that the symmetry which reverses the direction of motion
on the sphere $\omega\to-\omega$ re-arranges the fermions $s\to1-s$
. This appendix looks at the same idea in $AdS_{4}\times CP^{3}$,
as a check on our understanding of the formalism. 

\newcommand{\mydynkinCP}{
\rlap{\hspace{-3mm}
\begin{tikzpicture}[scale=0.56, darkred]
\draw (0,0) -- (3,0);
\draw (3,-0.5mm) -- (4.5,-2mm);
\draw (3,+0.5mm) -- (6,+4mm);
\draw [fill=white] (0,0) circle (2mm); 
\draw [fill=white] (1.5,0) circle (2mm); 
\draw [fill=white] (3,0) circle (2mm); 
\draw [fill=white] (4.5,-2mm) circle (2mm); 
\draw [fill=white] (6,+4mm) circle (2mm);
\draw (-2mm,0) -- (2mm,0);
\draw (0,-2mm) -- (0,2mm);
\draw (28mm,0) -- (32mm,0);
\draw (3,-2mm) -- (3,2mm);
\end{tikzpicture}
}}\newcommand{\smalldynkinCP}[5]{
\raisebox{-1mm}{
\begin{tikzpicture}[scale=0.6,semithick]

\draw (1,0) -- (3,0);
\draw (3,-0.5mm) -- (4,-1mm);
\draw (3,+0.5mm) -- (5,+3mm);

\draw [fill=#1] (1,0) circle (2mm); 
\draw [fill=#2] (2,0) circle (2mm); 
\draw [fill=#3] (3,0) circle (2mm); 
\draw [fill=#4] (4,-1mm) circle (2mm); 
\draw [fill=#5] (5,+3mm) circle (2mm);

\draw [thin] (8mm,0) -- (12mm,0);
\draw [thin] (1,-2mm) -- (1,2mm);

\draw [thin] (28mm,0) -- (32mm,0);
\draw [thin] (3,-2mm) -- (3,2mm);

\end{tikzpicture}
}
} 

We follow here the conventions of \cite{Zarembo:2010yz},%
\footnote{Note that \cite{Zarembo:2010yz} uses a different grading for the
superalgebra to \cite{Gromov:2008bz}. This should not matter for
classical strings, but does make figure 2 there look a little different
from table \ref{tab:List-of-modes-CP3} here. %
} in which the Cartan matrix is 
\[
A=\left[\begin{array}{ccccc}
0 & 1\\
1 & -2 & 1\\
 & 1 & 0 & -1 & -1\\
 &  & -1 & 2\\
 &  & -1 &  & 2
\end{array}\right]
\]
and the generators in weight space $\Lambda_{\ell}\in\mathbb{R}^{2,3}$
are: 
\[
\begin{array}{cccccc|cr}
\Lambda_{1} & \Lambda_{2} & \Lambda_{3} & \Lambda_{4} & \Lambda_{5} & \\
\mydynkinCP &  &  &  &  & \\
1 &  &  &  &  &  & B_{1} & i=3\\
\hline 1 & -1 &  &  & \phantom{-1} &  & F_{1} & 1\\
 & 1 & -1 &  &  &  & F_{2} & 2\\
\hline  &  & -1 & 1 & 1 &  & B_{2} & 4\\
\phantom{-1} &  &  & -1 & 1 &  & B_{3} & 5\rlap{.}
\end{array}
\]
We think of this as the matrix $B_{i\ell}=(\Lambda_{\ell})_{i}$ such
that $q_{i}=B_{i\ell}p_{\ell}$, where $p_{\ell}$ are the quasimomenta
corresponding to Cartan generators $\Lambda_{\ell}$ i.e. to nodes
of the Dynkin diagram, and $q_{i}$ are the quasimomenta with manifest
$OSp(2,2|6)$ symmetry as in \cite{Gromov:2008bz}. The lower five
of these are  defined $q_{11-i}=-q_{i}$, and $i=3,4,\ldots8$ describe
$CP^{3}$ while $i=1,2,9,10$ describe $AdS_{4}$. . Inversion symmetry
$p(\tfrac{1}{x})=Sp(x)$ is simple in terms of the $q_{i}$:%
\footnote{The inversion symmetry matrix for $p_{\ell}$ is 
\[
S=\begin{bmatrix} &  & 1 & -1 & -1\\
 & 1 &  & -1 & -1\\
1 &  &  & -1 & -1\\
 &  &  & 0 & -1\\
 &  &  & -1 & 0
\end{bmatrix}.
\]
} 
\[
q_{1}(x)=-q_{2}(\tfrac{1}{x}),\qquad q_{3}(x)=-q_{4}(\tfrac{1}{x}),\qquad q_{5}(x)=-q_{5}(\tfrac{1}{x}).
\]

Zarembo \cite{Zarembo:2010yz} gives two vacua, the only solutions
of $\vec{\kappa}A\,\vec{\kappa}=0$ and $S\,\vec{\kappa}=-\vec{\kappa}$,
and discards the second of these: 
\begin{align}
p_{\ell}(x) & =\frac{\Delta}{2g}\frac{x}{x^{2}-1}\kappa_{\ell},\quad\vec{\kappa}=(1,0,-1,0,0) & \Rightarrow\,\quad q(x) & =\frac{\Delta}{2g}\frac{x}{x^{2}-1}(1,1\:\vert\,1,1,0)\label{eq:vacua-pq-CP3}\\
p_{\ell}'(x) & =\frac{\Delta}{2g}\frac{x}{x^{2}-1}\kappa'_{\ell},\quad\vec{\kappa}^{\prime}=-(1,2,3,2,2) & \Rightarrow\quad q'(x) & =\frac{\Delta}{2g}\frac{x}{x^{2}-1}(1,1\:\vert\,-1,-1,0).\nonumber 
\end{align}
Written in weight space, it is clear that the second one differs by
a minus in the $CP^{3}$ sheets.%
\footnote{Here and in \eqref{eq:reversal-CP3} we make a choice over how to
treat $q_{5}$. However $q_{5}\to-q_{5}=q_{6}$ is already a discrete
symmetry of the algebraic curve, which re-organises the light modes
$(4,5)\to(4,6)$ etc, and changes the sign of angular momentum $J_{3}$
only. The choice to include this in $q\to q'$ results in this changing
the sign of every $CP^{3}$ charge which seems tidier. %
} It will thus differ by a minus in all its $SU(4)$ charges. But physically
we expect this to be little different; there is nothing sacred about
the direction of motion of the initial BMN particle.

The same symmetry will exist for an arbitrary solution: given some
classical curve $q_{i}(x)$, the reversed curve 
\begin{equation}
q_{i}'(x)=\begin{cases}
q_{i}(x), & i=1,2\mbox{ (and thus }9,10)\\
-q_{i}(x), & 3\leq i\leq8
\end{cases}\label{eq:reversal-CP3}
\end{equation}
is also a valid algebraic curve. This will have charges $J'=-J$,
$Q'=-Q$ but should be physically equivalent. In particular the frequencies
of its vibrational modes should be identical, but calculating these
according to the usual rules \cite{Gromov:2008bz} does not give the
same answer. For instance, taking the example of the vacuum solution
again, the heavy fermion $(1,7)$ has become massless, unlike any
of the modes of the original solution:
\begin{align*}
2\pi n{}_{(1,7)} & =q{}_{1}-q{}_{7}=2\frac{\Delta}{2g}\frac{x}{x^{2}-1}\\
2\pi n'_{(1,7)} & =q'_{1}-q'_{7}=0,\qquad2\pi n'{}_{(1,4)}=q'_{1}+q'_{7}=2\frac{\Delta}{2g}\frac{x}{x^{2}-1}.
\end{align*}
What we should obviously do is to insert the same minus in the definition
of the modes: if we include a $(1,4)$ mode for the primed quasimomenta
(i.e. allow cuts connecting sheet $q'_{1}$ not to $q'_{7}=-q'_{4}$
but to $+q'_{4}$) then we will recover the same frequency as the
$(1,7)$ mode on the unprimed quasimomenta. 

\begin{table}
\begin{tabular}{lccccc}
\hline 
Light Bosons: & $(4,5)$ & \smalldynkinCP{white}{white}{white}{\colourone}{white} & $\circlearrowright-$ &  & \tabularnewline
 & $(4,6)$ & \smalldynkinCP{white}{white}{white}{white}{\colourone} & $\circlearrowright-$ &  & \tabularnewline
 & $(3,5)$ & \smalldynkinCP{\colourone}{\colourone}{\colourone}{\colourone}{white} & $\circlearrowright-$ &  & \tabularnewline
 & $(3,6)$ & \smalldynkinCP{\colourone}{\colourone}{\colourone}{white}{\colourone} & $\circlearrowright-$ &  & \tabularnewline
\hline 
Light Fermions: & $(1,5)$ & \smalldynkinCP{white}{\colourone}{\colourone}{\colourone}{white} & $\leftrightarrow$ & $(1,6)$ & \smalldynkinCP{white}{\colourone}{\colourone}{white}{\colourone}\tabularnewline
 & $(2,5)$ & \smalldynkinCP{white}{white}{\colourone}{\colourone}{white} & $\leftrightarrow$ & $(2,6)$ & \smalldynkinCP{white}{white}{\colourone}{white}{\colourone}\tabularnewline
\hline 
Heavy Bosons, $CP$: & $(3,7)$ & \smalldynkinCP{\colourone}{\colourone}{\colourone}{\colourone}{\colourone} & $\circlearrowright-$ &  & \tabularnewline
$\phantom{.}$\hfill$AdS$: & $(1,10)$ & \smalldynkinCP{white}{\colourtwo}{\colourtwo}{\colourone}{\colourone} & $\circlearrowright$ &  & \tabularnewline
 & $(1,9)$ & \smalldynkinCP{white}{\colourone}{\colourtwo}{\colourone}{\colourone} & $\circlearrowright$ &  & \tabularnewline
 & $(2,9)$ & \smalldynkinCP{white}{white}{\colourtwo}{\colourone}{\colourone} & $\circlearrowright$ &  & \tabularnewline
\hline 
Heavy Fermions: & $(1,7)$ & \smalldynkinCP{white}{\colourone}{\colourone}{\colourone}{\colourone} & $\to$ & $(1,4)$ {*} & \smalldynkinCP{white}{\colourone}{\colourone}{white}{white}\tabularnewline
 & $(2,7)$ & \smalldynkinCP{white}{white}{\colourone}{\colourone}{\colourone} & $\to$ & $(2,4)$ {*} & \smalldynkinCP{white}{white}{\colourone}{white}{white}\tabularnewline
 & $(1,8)$ & \smalldynkinCP{\colourone}{\colourtwo}{\colourtwo}{\colourone}{\colourone} & $\to$ & $(3,1)$ {*} & \smalldynkinCP{\colourone}{white}{white}{white}{white}\tabularnewline
 & $(2,8)$ & \smalldynkinCP{\colourone}{\colourone}{\colourtwo}{\colourone}{\colourone} & $\to$ & $(3,2)$ {*} & \smalldynkinCP{\colourone}{\colourone}{white}{white}{white}\tabularnewline
\hline 
\end{tabular}

\protect\caption[Fake caption without tikz figures.]{List of modes in the $AdS_{4}\times CP^{3}$ algebraic curve, with
$\to$ indicating the effect of reversing the direction \eqref{eq:reversal-CP3}.
A star marks modes which did not exist before. As before the colouring
of the nodes is $-k_{\ell\,(i,j)}$ with $\smallcirc{\colourone}= 1$
and $\smallcirc{\colourtwo}=2$, and $2\pi n=-k_{\ell\,(i,j)}A_{\ell m}p_{m}=q_{i}-q_{j}$.
 \label{tab:List-of-modes-CP3}}
\end{table}

We can do this for all the modes, and translating back to descriptions
in terms of $p_{\ell}$ we get the map shown in table \ref{tab:List-of-modes-CP3}.
The light modes just get re-arranged, and the heavy bosons are left
alone. This map gives new modes only for the heavy fermions, marked
with a star. 
\begin{itemize}
\item Note that the list of allowed modes is extra information not encoded
in the finite gap equations, by which in this case we mean (3.26)
of \cite{Zarembo:2010yz}. Each mode is some correlated set of densities
$\rho_{\ell}(y)$. This is true of both the original set of modes,
from \cite{Gromov:2008bz}, and the new set for the reversed BMN solution. 
\item Notice that these new modes don't excite either of the momentum-carrying
nodes $\ell=4,5$, while all the old light modes excited one, and
all the old heavy modes excited both. However our notion of which
nodes are momentum-carrying depends on the choice of vacuum, and with
the reversed BMN $\vec{\kappa}'$ it is not only the nodes $\ell=4,5$:\newcommand{\tempspace}{\hspace{4.5mm}}
\newcommand{\tempup}[1]{\smash{\raisebox{3mm}{\scriptsize{$#1$}}}} 
\begin{align}
P & =\kappa_{\ell}A_{\ell k}P_{k}=P_{4}+P_{5} &  & \underset{\hspace{2mm}\hspace{20mm}1\tempspace\tempup{1}}{\smalldynkinCP{white}{white}{white}{white}{white}}\nonumber \\
P' & =\kappa'_{\ell}A_{\ell k}P_{k}=-2P_{1}+2P_{3}-P_{4}-P_{5} &  & \underset{\hspace{2mm}-2\tempspace\,\tempspace2\tempspace-1\tempspace\tempup{-1}}{\smalldynkinCP{white}{white}{white}{white}{white}}\label{eq:mom-CP3}
\end{align}
Counted according to $P'$, the new heavy modes do all excite two
momentum-carrying nodes.%
\footnote{The light modes appear to excite $\pm1$, since table \ref{tab:List-of-modes-CP3}
is not careful about the overall sign, i.e. the sign of the mode number.
The modes which need a minus are marked $\circlearrowright-$. This
is also the reason that $(3,1)$ is not written $(1,3)$. %
}
\item In the worldsheet string theory, $CP^{3}$ is described by $Z\in\mathbb{C}^{4}$
with $Z\sim\lambda Z$ for any $\lambda\in\mathbb{C}$. The effect
of \eqref{eq:reversal-CP3} here is to conjugate each embedding co-ordinate
$Z_{i}'=Z_{i}^{*}$, thus reversing all $CP^{3}$ angular momenta.
This is clearly a symmetry of the action. 
\end{itemize}
These features are the same as for the $AdS_{3}\times S^{3}\times S^{3}\times S^{1}$
case: there too the ``$0f$'' mode is obtained from the heavy fermion
``$4f$'', and does not excite either of what were initially thought
of as momentum-carrying nodes. The crucial difference is that in $AdS_{4}$
this is just some discrete symmetry, which we could avoid thinking
about by always rotating our co-ordinates to put the particle's momentum
into a standard direction before working out the monodromy matrix.
If someone else made a different choice there is no surprise that
they end up using a slightly different formalism.%
\footnote{Another way to look at the change in \eqref{eq:vacua-pq-CP3} is as
exchanging in $SU(4)$ the highest-weight state $(1,1,0)$ with the
lowest-weight one $(-1,-1,0)$ . We thank Olof Ohlsson Sax for pointing
this out. %
} In $AdS_{3}$ on the other hand, there is a continuous set of physically
distinct (non-BPS) solutions connecting the two opposite directions,
and the same formalism ought to cover all of them, or at least both
ends. We only understood $6/8$ of this, at either end, but the connection
enabled us to fill in the remaining $2/8$ of the whole. 

In this regard $AdS_{5}\times S^{5}$ will be just like $AdS_{4}\times CP^{3}$:
we can write down the modes for an alternative vacuum, the reversed
BMN solution, but there is never a need to do so. 

\bibliographystyle{my-JHEP-4}
\bibliography{
complete-library-processed-plus}

\end{document}